\documentclass[aps,twocolumn,secnumarabic,nobalancelastpage,amsmath,amssymb,nofootinbib]{revtex4}

\usepackage{subfigure}
\usepackage{graphicx}
\usepackage{url}
\usepackage{natbib}

\begin{document}

\title{Finding New Order in Biological Functions from the Network Structure of Gene Annotations}
\author{Kimberly Glass\,$^{1,2,3}$\footnote{contact: kglass@jimmy.harvard.edu}, and Michelle Girvan\,$^{3,4,5}$}
\address{$^{1}$Biostatistics and Computational Biology, Dana-Farber Cancer Institute, Boston, MA, USA\\
$^{2}$Department of Biostatistics, Harvard School of Public Health, Boston, MA, USA\\
$^{3}$Physics Department, University of Maryland, College Park, MD, USA\\
$^{4}$Institute for Physical Science and Technology, University of Maryland, College Park, MD, USA\\
$^{5}$Santa Fe Institute, Santa Fe, NM, USA}

\begin{abstract}
The Gene Ontology (GO) provides biologists with a controlled terminology that describes how genes are associated with functions and how functional terms are related to each other.  These term-term relationships encode how scientists conceive the organization of biological functions, and they take the form of a directed acyclic graph (DAG).  Here, we propose that the network structure of gene-term annotations made using GO can be employed to establish an alternate natural way to group the functional terms which is different from the hierarchical structure established in the GO DAG.  Instead of relying on an externally defined organization for biological functions, our method connects biological functions together if they are performed by the same genes, as indicated in a compendium of gene annotation data from numerous different experiments.  We show that grouping terms by this alternate scheme is distinct from term relationships defined in the ontological structure and provides a new framework with which to describe and predict the functions of experimentally identified sets of genes.  Tools and supplemental material related to this work can be found online at http://www.networks.umd.edu.
\end{abstract}

\maketitle

\section{Introduction}

The Gene Ontology (GO) \cite{citeulike:6574187}\cite{citeulike:1049349} has been around for over a decade, during which time it has been widely utilized both to validate and to predict the results of biological experiments (see, for example \cite{citeulike:1530640,citeulike:7237554, citeulike:2817462, citeulike:7367164, citeulike:134628, citeulike:624114, citeulike:5400242}).  The structure of the ontology, where different ``categories'' or terms are related to each other in a hierarchical fashion, provides a well-established format with which to classify and subclassify all biological functions and processes.  This classification approach is well-structured and well-characterized, however, we seek to determine if it is the only natural way in which to classify this type of biological information.  We address two main questions.  First, does there exist another natural way to organize the functional terms that is distinct from the ontological organization? Secondly, if such an alternate classification exists, can it be used to interpret biological data?

In recent years, complex networks tools have been used alongside traditional bioinformatics techniques to study many different kinds of biological networks \cite{newman_siam_review}, including, but not limited to, gene regulatory networks \cite{citeulike:101, citeulike:2855605}, protein-protein interaction networks \cite{citeulike:478707, wagner01a}, and metabolic networks \cite{Guimera05, citeulike:2648743}.  Developments in network theory provide the computational tools needed to calculate global properties of such networks, lending insights into the behavior of the systems represented by these networks.  For example, many networks exhibit community structure, meaning that there are clusters of nodes in the network within which edges are relatively dense \cite{citeulike:81501}.  Within the field of complex networks, many recent research papers \cite{citeulike:4091170, citeulike:6574280,86, citeulike:1837596} have focused on the development of methods to detect such module in various types of networks \cite{86}\cite{citeulike:1837596} in a computationally efficient and accurate manner \cite{citeulike:95936}.  In this study, we leverage the community structure in gene annotation networks to develop an endogenous organization of biological functions.

Ontologies are utilized across many disciplines including economics, artificial intelligence, engineering, library science, and biomedical informatics (for example, \cite{EcomomicWorldView,citeulike:1735454,citeulike:4149292}).  The Gene Ontology, specifically, describes the relationships between different biological concepts or functions \cite{citeulike:6574187}.  It breaks these concepts into three main domains, or distinct ontologies: ``Biological Process'' (BP), describing sets of molecular events, ``Molecular Function'' (MF), describing the activities of gene products, and ``Cellular Component'' (CC), describing parts of a cell or its external environment.  Each of the three primary domains in GO takes the form of a directed acyclic graph (DAG), in which ``child'' functional categories, or ``terms'', are subclassified under one or more ``parent'' terms.  Each parent and all its subsequent progeny therefore define multiple, overlapping, sets of terms, or ``branches'' in GO.  Using GO, genes are annotated to individual terms representing their particular role in a cell, and these annotations are transitive up the relationships in the DAG such that each ``parent'' term takes on all the gene annotations associated with any of its progeny \cite{GO2001}.  In the past there has only been minimal investigation of how biological functions might be related to each other outside of the ontology structure, with the majority focusing on discovering individual links between functions \cite{citeulike:2817462} rather than investigating the structure as a whole.  In this work, we propose an alternate classification of functional terms that relies on gene-term annotations rather than ontological relationships.

\begin{figure}
\label{Approach}
\includegraphics[width=220px]{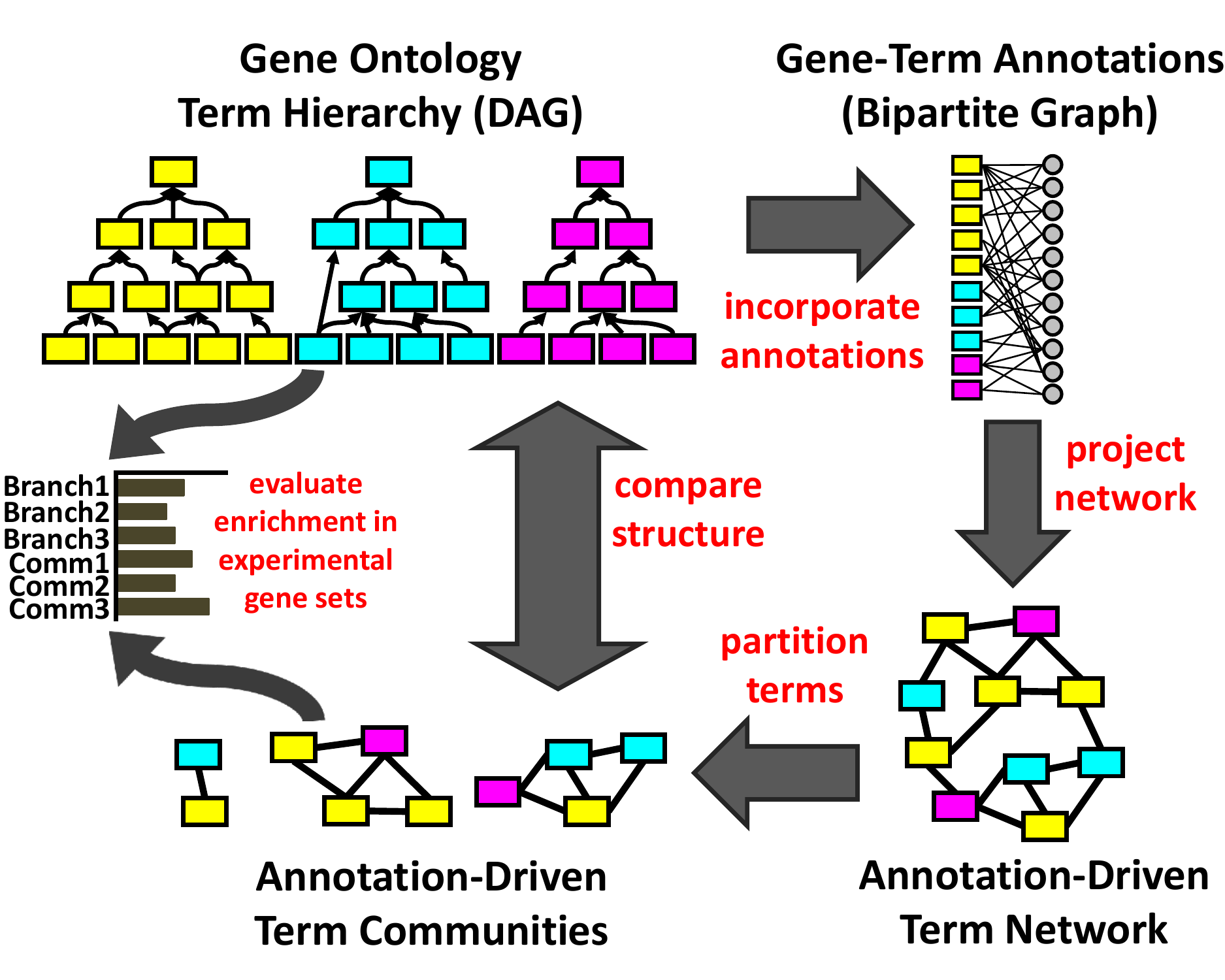}
\caption{Visual representation of our approach.  First, we summarize gene annotations made to functional terms in the Gene Ontology hierarchy as a gene-term bipartite graph.  From these gene-term relationships, we project a term-term network.  We partition this network into communities and compare those term communities to branches of terms in the DAG.  Finally, we perform functional enrichment analysis on experimentally-defined gene sets using both the term communities and GO branches.}
\end{figure}

Our complex networks approach to organizing biological functions using annotations made to the Gene Ontology is outlined in Figure \ref{Approach}.  We begin by considering term relationships defined by the GO hierarchy.  We then add in gene-term annotation information collected from numerous different experiments and encapsulate these connections in the form of a bipartite network.  Next, we use this bipartite network to construct another network describing the relationships between functional terms based on shared gene annotations.  We apply community structure finding algorithms to partition this annotation-driven network into communities of terms and compare these communities to branches (ontological groupings of terms) from the GO hierarchy.  We show that, although there are some similarities, there are also very strong differences between the two ways of organizing terms.  Finally, we test the applicability of the community-derived classification, utilizing functional analysis techniques to evaluate the enrichment of cancer signatures (sets of genes associated with cancer) in both term communities and GO branches. We find that certain signatures are enriched primarily in our term communities and not GO branches.  Therefore, we suggest that by linking functional terms based on shared genes, we can create an alternate, biologically meaningful, network-derived organization of terms that is both distinct from the GO DAG and can be used to investigate biological systems.  Tools and supplemental material related to this work can be found online at http://www.networks.umd.edu.

\section{Methods}

\subsection{Characterizing Gene Ontology Annotations in a Bipartite Graph}
In the following analysis we explore if there exists an alternate, natural way to classify terms that is distinct from the ontology structure.  To begin, we use term-term ontology relationships and gene-term annotation information for human genes downloaded from the GO website (geneontology.org) to construct a gene-term bipartite network.  We choose to represent this network in the form of an $n_G \times n_T$ adjacency matrix, where $n_G$ is the total number of genes and $n_T$ is the total number of terms listed in the annotation file.  In this matrix a value of one indicates a known connection between the corresponding gene and term, and a value of zero indicates that the gene is not associated with that term.  Thus,
\begin{equation}
B_{pi}=
\begin{cases}
1&\text{if gene $p$ is annotated to term $i$}\\
0&\text{if gene $p$ is not annotated to term $i$}
\end{cases}.
\label{Bqi}
\end{equation}
This bipartite network represents a summary of the relationships between $18930$ genes and $15033$ functional terms, derived from many different types of biological evidence and contributed to by multiple laboratories\cite{UniProtGOA}.  We note that although GO is broken into three primary domains and gene-annotations are made to the ontology for many species, for simplicity in the following analysis we will combine information from all three domains and use annotation information only that pertains to human genes.  Domain-specific and comparative species analysis is provided in the Supplemental Material.

\subsection{Constructing a Term Network from Gene Ontology Annotations}
\label{ProjectNetworks}

Next, we used gene-term annotations to construct a network representing term-term relationships.  Using the bipartite network (Equation \ref{Bqi}) one could create a term network by simply joining together any pair of terms that share common genes; however, the number of genes annotated to each term has a heavy-tailed distribution (see Supplemental Figure S\ref{BranchCommSizeDist}), thus this approach would lose a large amount of information as connections between pairs of highly-annotated terms would be given the same weight as connections between pairs of terms with only a few annotations.  We correct for the skewed term degree distribution by constructing a diagonal weighting matrix, $w$, and then projecting a term network $T$, whose edges are modified by this weighting matrix:
\begin{equation}
w_{ij}=\frac{\delta_{ij}}{\displaystyle\sum_{q=1}^{n_G}B_{qi}}, \qquad T=w'B'Bw.
\label{weightingmatrix}
\end{equation}
The values of $T_{ij}$ take a maximum value of one when terms $i$ and $j$ each only have the same single gene annotation and a minimum value of zero when none of the genes annotated to term $i$ are annotated to term $j$.  The use of the weighting matrix emphasizes the weights of network edges between low degree terms.  Since these terms represent biological functions performed by only a handful of genes, we believe this weighting is more likely to capture highly-specific shared biological information.

\subsection{Identifying Communities of GO terms}
\label{PartitionNetwork}

Finally, we seek to identify the community structure in annotation-driven term-term relationships, or clusters of terms within which there are many or high-weight relationships in our projected network (Equation \ref{weightingmatrix}), but between which there are only few or low-weight relationships.  In order to quantify the strength of community structure we use a quantity known as modularity \cite{citeulike:6574280}.  Modularity ($Q$) can be defined as:
\begin{equation}
Q=\frac{1}{2m}\displaystyle\sum_{ij}\left[A_{ij}-\left(1+\frac{r}{\langle k\rangle}\right)\frac{k_ik_j}{2m}\right]\delta(x_i,x_j)
\label{ResolutionModularity}
\end{equation}
where $\delta$ is the Kronecker delta function, $x_i$ is the community of node $i$, $k_i$ is the degree of node $i$, $A$ is the adjacency matrix, a matrix with values representing the weight between nodes $i$ and $j$, and $m$ is the total weight of the edges in the network\cite{86,Arenas2008}.  Traditionally, in order to partition a network into communities, the resolution parameter, $r$ in Equation \ref{ResolutionModularity}, is set equal to zero.  Varying this value allows one to look for alternate divisions of a network into communities at different scales, or resolutions, with $r>0$ uncovering sub-structures in the network \cite{Arenas2008}.

\begin{figure*}
\includegraphics[width=500px]{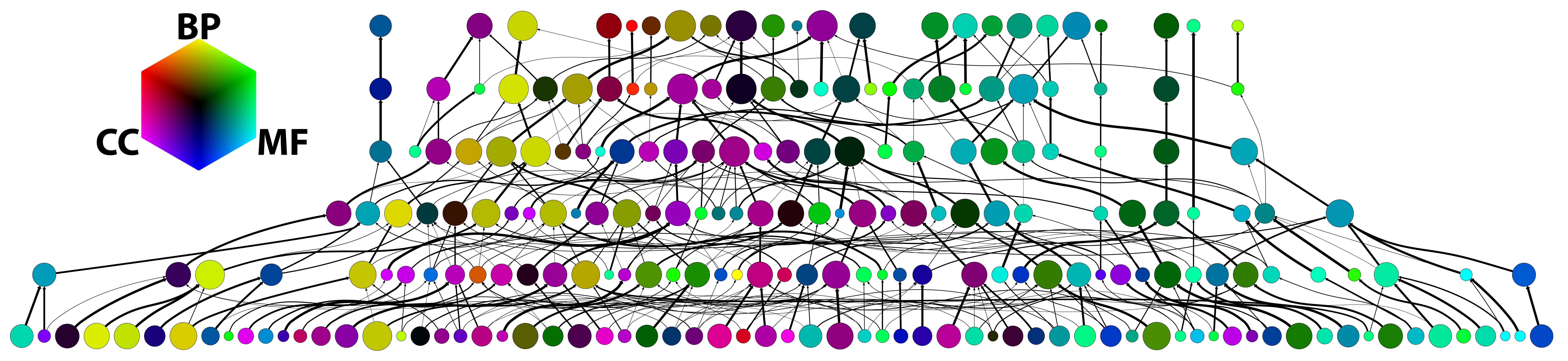}
\caption{Visualization of communities (circles) of GO terms found at the six lowest levels of resolution (rows), in increasing order (top to bottom).  The width of the line connecting two communities is proportional to the percentage of terms in the child community that are also in the parent community.  The size of communities is proportional to the log of the number of terms in the community.  Color represents the normalized percentage of terms in the community which belong to the BP (yellow), MF (cyan) and CC (magenta) primary domains.}
\label{CommunityDAG}
\end{figure*}

We used a weighted version of the Fast Greedy Community Structure algorithm \cite{citeulike:95936} to investigate the community structure of our term network, and found fifty-six communities at maximum modularity.  We then implemented a modified version of the Fast Greedy that maximizes modularity for non-zero values of the resolution parameter in order to find many different viable partitions.  We varied the resolution parameter several orders of magnitude and found $11491$ different communities (see Supplemental Table \ref{Table1}).  We gave our communities numeric identities that vary from TC:0000001 to TC:0011491 and will refer to them as such in the following analysis.  The different values of the resolution parameter were chosen to give community sizes that were roughly similar to those defined by the branches at different levels of the GO DAG (see Supplemental Figure S\ref{BranchCommSizeDist}).  Like GO branches, which represent overlapping sets of functional categories rather than one discreet partition of terms, communities found at different resolutions are highly overlapping and represent functional structure at many different levels of specificity.

\section{Results: An alternate ``Natural'' Grouping of GO Terms}

\subsection{Term Communities and GO Branches Represent Distinct Collections of Biological Functions}

To better understand the relationships between the communities found at different resolutions, we visualized the term communities with ten or more members for the six lowest values of resolution used (Figure \ref{CommunityDAG}).  In this visualization each community is represented by a single circle, whose radius scales as the log of the number of terms belonging to that community and whose color corresponds to the percentage of members from each primary domain that belong to that community.  Between the communities found at adjacent resolutions, we draw a line from a community at a higher resolution to a community at a lower resolution if at least 10\% of the members of the community from the higher resolution also belong to the community at the lower resolution.  The thickness of the line is indicative of the overlap between the two communities.  For more details on the visualization process see the Supplemental Material.

\begin{figure}
\makeatletter
\def\@captype{figure}
\makeatother
\begin{center}
\subfigure[Distribution of $J_m$ for Branches and Communities]{\includegraphics[width=200px]{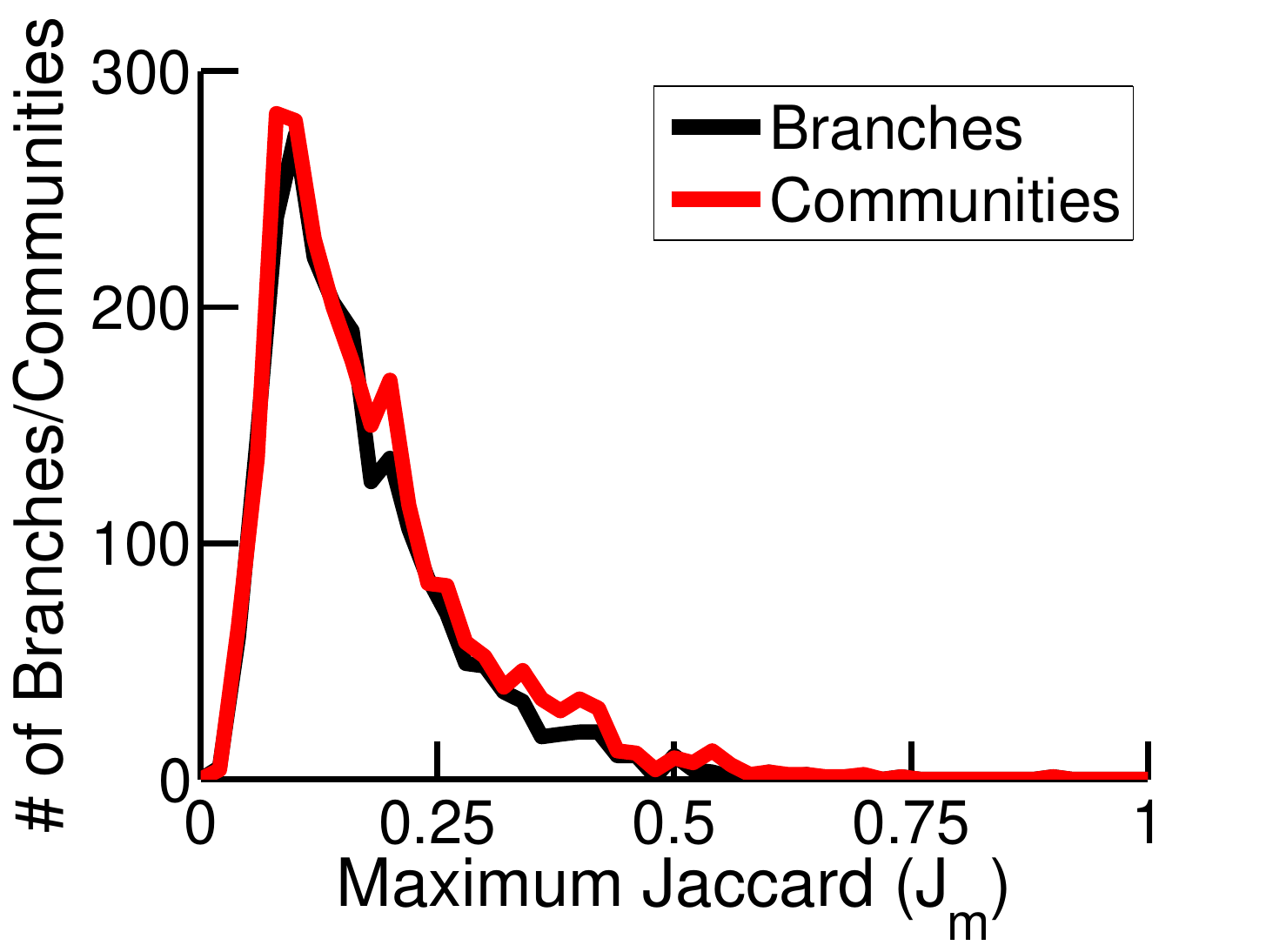}\label{BranchCommJaccardDist}}
\subfigure[TC:0007391, $J_m=0.6667$ with GO:0070570]{\includegraphics[width=90px]{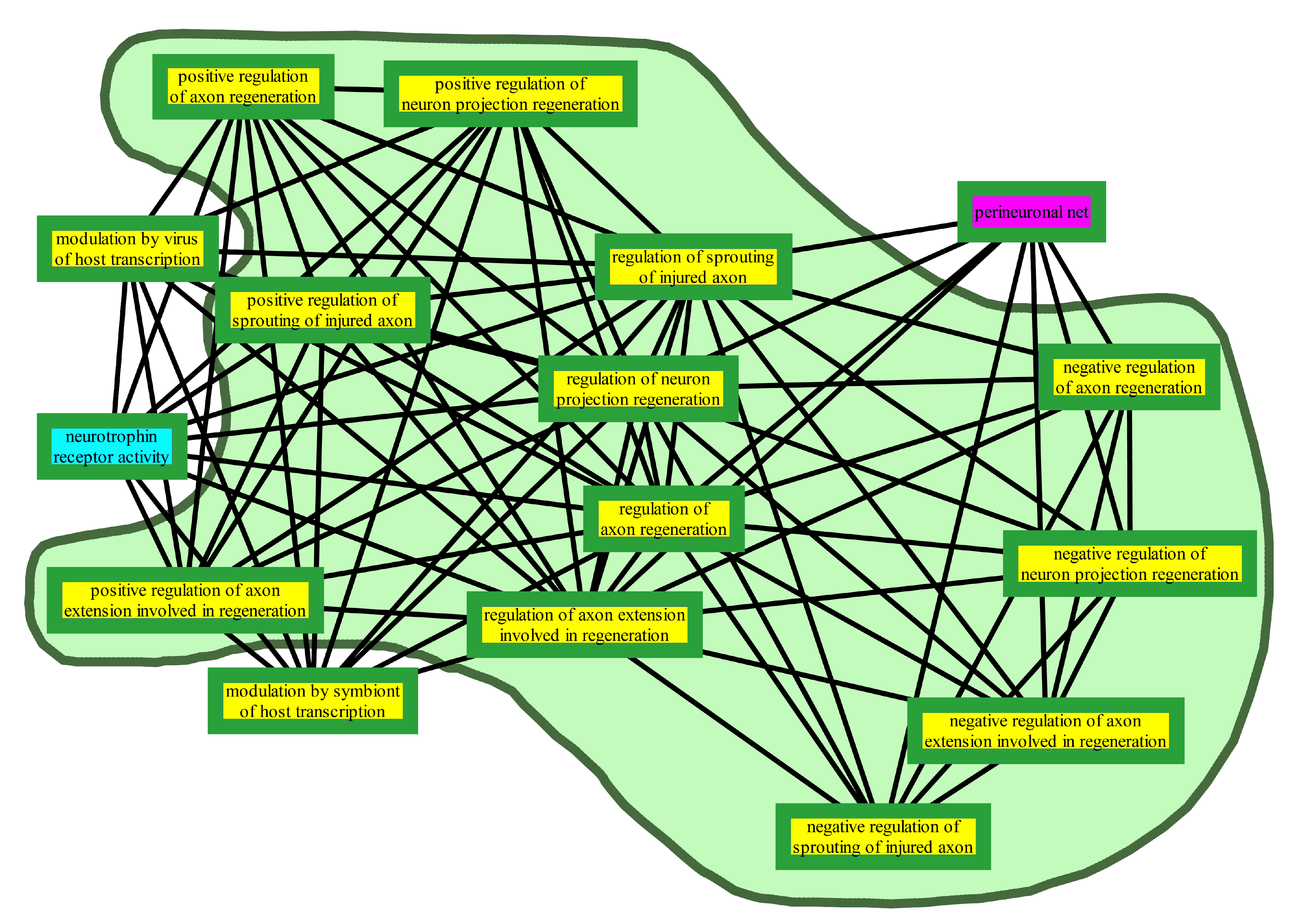}\hspace{10px}\includegraphics[width=90px]{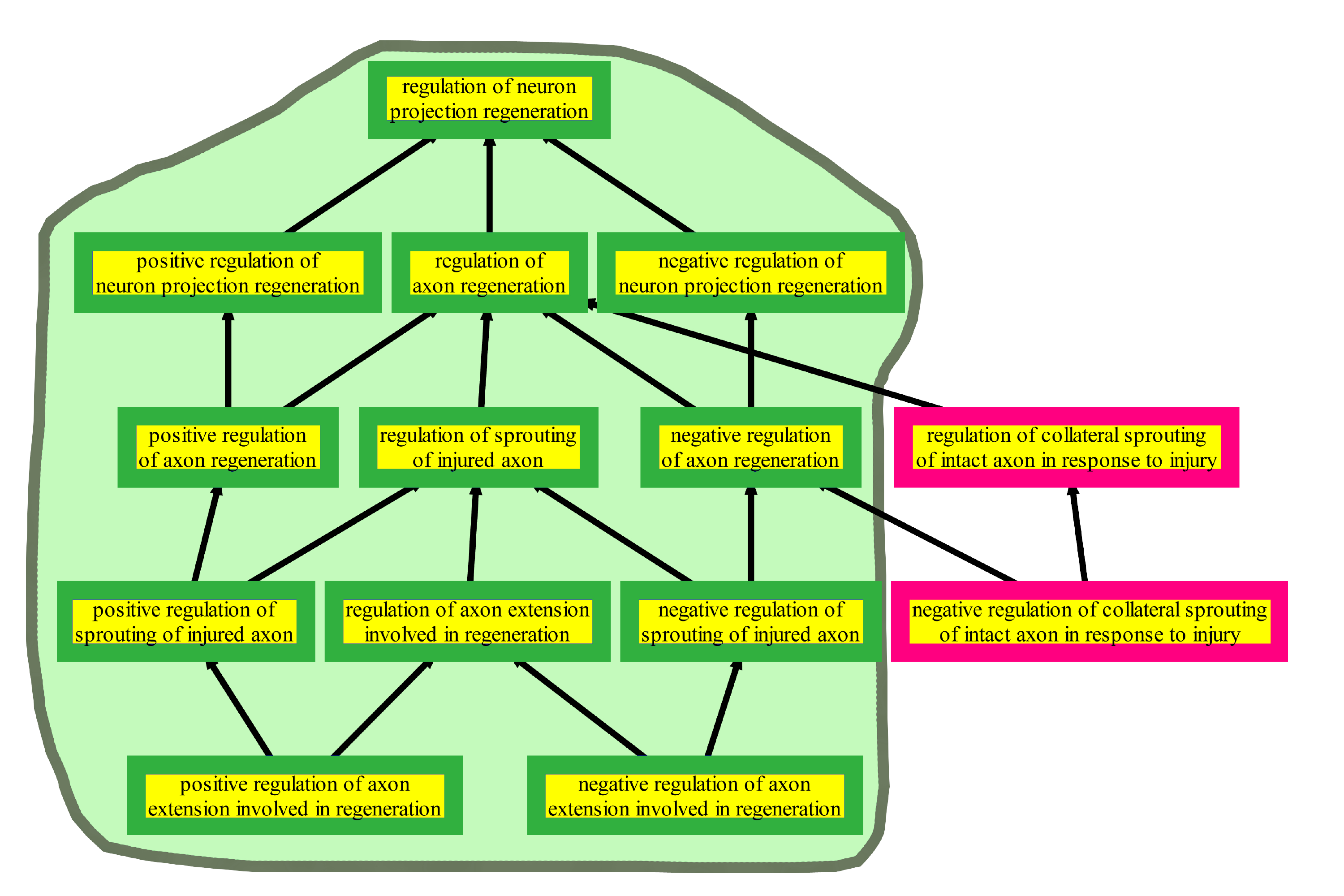}\label{Comm7391vsBranch14090}}
\subfigure[TC:0000936, $J_m=0.1667$ with GO:0060538]{\includegraphics[width=90px]{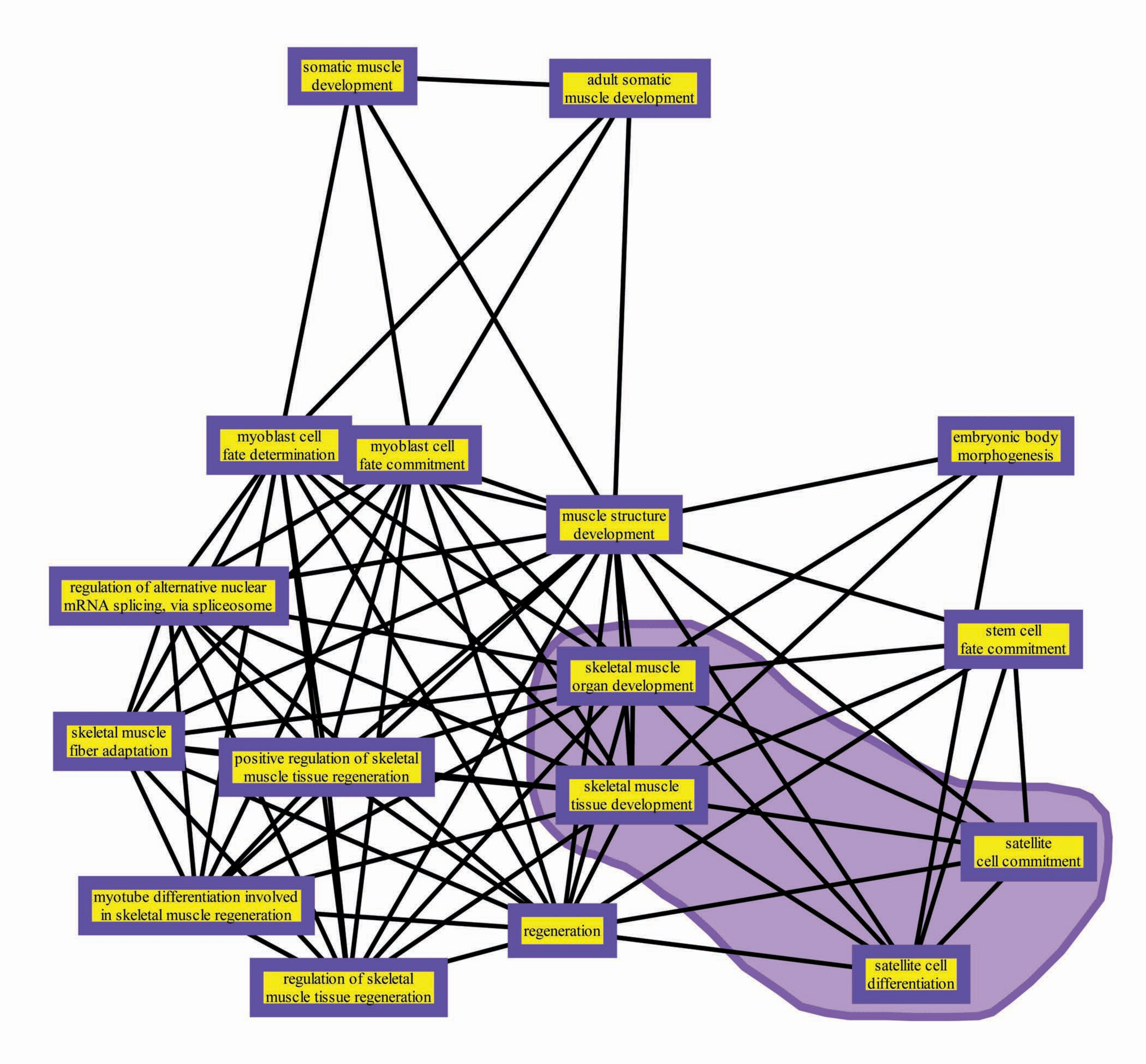}\hspace{10px}\includegraphics[width=120px]{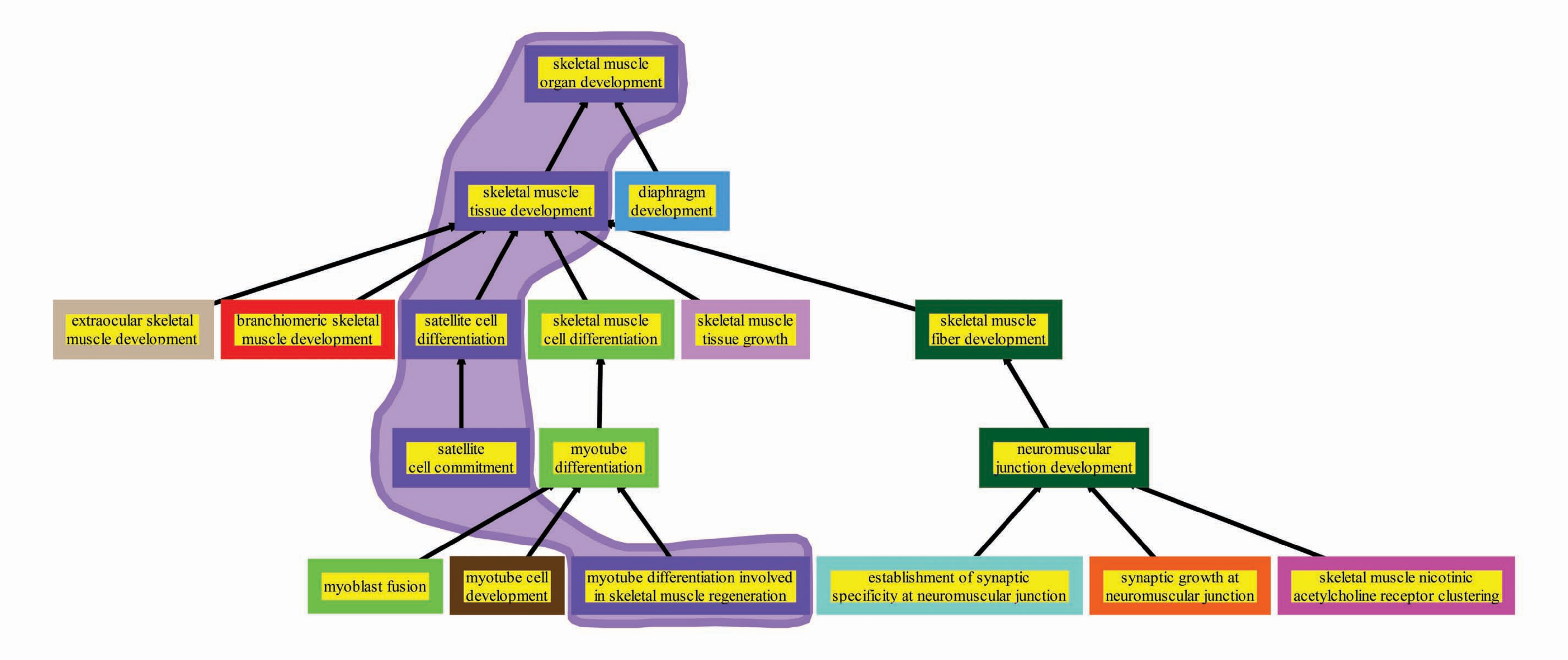}\label{Comm936vsBranch10413}}
\end{center}
\caption{A comparison of branches in the GO DAG and term communities found by partitioning the term network.  (A) Distribution of $J_m$, the maximum similarity a community or branch with ten or more members has compared to all other branches or communities with ten or more members, respectively.  Although a small number of communities and branches have similar memberships, most are highly dissimilar. (B)-(C) Two example comparisons between communities and branches: (B) TC:0007391 compared to GO:0070570, and (C) TC:0000936 compared to GO:0060538.  In each panel on the left hand side a community and its inter-community connections in the annotation-driven term network is shown and on the right hand side the branch with which that community has the the highest Jaccard similarity is illustrated.  In the right panel edges represent the ontological associations defined by the Gene Ontology term hierarchy. Each term member of the community or branch is colored both by its associated primary domain (inner color - BP:yellow, MF:cyan, CC:magenta) and its community membership (outer color), determined at the same resolution value as the illustrated community.  Terms common between each community and branch pair are circled.}
\label{MaxJaccard}
\end{figure}

The structure of annotation-driven term relationships is distinct from the structure of those relationships as defined by GO branches.  This is evidenced clearly by the fact that, although each GO branch can only belong to one primary ontology, and thus would be pure yellow, cyan or magenta in this type of visualization, communities, even smaller ones and those found at higher resolutions, generally contain members from multiple ontologies, resulting in a rainbow of colors.  We also observe that communities at higher resolutions do not merely represent the ``splitting apart'' of communities at lower resolutions (represented by a child community only connecting to a single parent), but instead each resolution often brings about a new way of partitioning the network.  An analogous visualization of GO branches reveals a similar a complex partitioning of terms in the GO DAG, albeit segregated by primary domain (see Supplemental Figure \ref{VizBranches}).

Next we directly compared the membership of the term communities with that of branches in the GO DAG.  In order to quantify the similarity between each community and branch, we calculated the Jaccard similarity, which takes the value $J(x,y)={| x \cap y |}/{|x \cup y |}$.  Then, for each community ($x$), we determined the corresponding branch ($y$) that has the highest overlap in membership by this measure: $J_m(x)= \max \{J(x,y):y\in Y\}$, and vice versa.  Because the exact value of the Jaccard similarity is highly sensitive to incremental changes in set membership when comparing sets with only a few members, we will limit all the following analysis to communities and branches that contain ten or more terms in order to focus on the most robust results.  Figure \ref{BranchCommJaccardDist} shows the distribution of $J_m$ comparing these 2370 communities and 2151 branches.  Although a handful of communities and branches are quite similar to each other, the majority of communities are dissimilar to the GO Branches and vice versus.  We have repeated this analysis constructing the term network and corresponding partitions three more times, using annotations specific to each of the three primary ontologies, and observe similar results (see Supplemental Figure S\ref{MaxJBP}-\ref{MaxJCC} and Supplemental Table \ref{Table2}).

To better interpret these values, we selected several communities to inspect more closely. First we selected a community with a very high $J_m$ value to inspect (Figure \ref{Comm7391vsBranch14090}).  TC:0007391 is most similar to GO:0070570 (``regulation of neuron projection regeneration'') with $J_m=0.6667$.  It is interesting that in addition to members from the BP domain, TC:0007391 also includes two members from the MF and CC domains, ``neutrophin receptor activity'' and ``perineuronal net'' respectively, the former of which is involved in the regeneration of injured axons \cite{citeulike:6926373} while the degradation of the latter has been shown to favor axon regeneration \cite{citeulike:11319900}.  This indicates that terms found in the community but not the branch are consistent with known biology.

Next we selected TC:0000936, which is most similar ($J_m=0.1667$) to GO:0060538 (Figure \ref{Comm936vsBranch10413}). We note that that the dissimilarity found between this community and branch cannot be attributed to community membership from multiple primary domains, as all of TC:0000936's members belong to the ``Biological Process'' primary domain.  Interestingly, the branch defined by GO:0060538 has members that belong to eleven distinct communities, demonstrating that not only are communities often distinct from branches, within the branches themselves the annotation-driven classification is often very distinct from the defined ontological relationships.  This pair is a representative example of the maximal shared information that is typically found between a community and branches, therefore we conclude that although there is occasional similarity between our found communities and GO branches, the communities are not simply a recapitulation of the DAG.

\subsection{Capturing the Biological Information in Term Communities}

We have illustrated that our term communities represent a natural partitioning of functional terms that is distinct from the GO DAG, however, the biological meaning of these communities is, at this point, unclear.  On a mathematical level they represent sets of biological functions that are generally performed by the same collection of genes.  Labeling and understanding the biological meaning behind these communities is vital if they are to have the same wide-range applications as the GO branches.  Therefore, in order to easily interpret the contents of an individual community we choose to summarize the descriptions of its member terms in the form of a word cloud, coloring each word in the cloud based on the normalized percentage of times the members is it derived from  belong to each primary domain, and scaling the size of each word by the statistical enrichment of its frequency in that community (for details see Supplemental Material).  We illustrate the biological content of two communities in Figure \ref{Comm400Cloud}-\ref{Comm61Cloud}.

The word clouds illustrate a richness of biological information in term communities.  Although Community TC:0000400 (Figure \ref{Comm400Cloud}) contains $335$ members harking from all three primary domains, the word cloud presentation easily summarizes this information.  The individual words are often contained in terms associated with multiple domains, resulting in a complex coloration, but reveal that this community includes biological concepts related to various types of RNA, including ``rRNA'', ``tRNA'', ``mRNA'', ``LSU-rRNA'', ``SSU-rRNA'', ``ncRNA'', ``RNA-polymerase'' and more.  In contrast, TC:0000061 contains many words related to the heart such as ``cardiac'', ``muscle'', ``ventricle'', ``ventricular'' and ``heart'' (Figure \ref{Comm61Cloud}).  Neither community is very similar to any particular branch in GO, although they represent similar biological information.  TC:0000400 is most similar ($J_m=0.22146$) to GO:0016070 or ``RNA metabolic process'', and TC:0000061 has the highest similarity ($J_m=0.149$) with GO:0072358, or ``cardiovascular system development.''

We point out that one can also represent the biological information contained in branches in the form of word clouds, although, because the members of each branch can only belong to one of the three primary domains, all the words in the cloud will be the same color.  Two branches are illustrated in Figure \ref{Branch1Cloud}-\ref{Branch743Cloud}.  The first, GO:0000003, or ``reproduction'' clearly contains terms pertaining to sex-related processes as it contains words such as ``female'', ``sex'', ``prostate'' and ``male.''  Similarly, the cloud for GO:0002376, whose parent term name is ``immune system process'' contains words pertaining to the immune system.

\subsection{Term Communities can be used to Evaluate and Predict Genetic Function}

Finally, we wanted to test how our communities might be used in one common application of the Gene Ontology: functional enrichment analysis.  To begin, we downloaded a collection of experimentally derived genes sets from the Gene Signatures Database (GeneSigDB) \cite{GeneSigDB2012}.  This database is a manual curation of previously published gene expression signatures, focusing primarily on cancer and stem cell signatures \cite{GeneSigDB2010}, and includes $509$ human signatures that contain at least $100$ and less than $1000$ genes annotated in the Gene Ontology.  To perform the functional enrichemnt analysis we use Annotation Enrichment Analysis \cite{citeulike:11319934} since it has been shown to better estimate the biological functions of experimentally derived sets of genes, and has a conceptual framework conducive to estimating functional enrichment between sets of terms and genes, rather than simply between two sets of genes (for more details see Supplemental Material).

\begin{figure}
\makeatletter
\def\@captype{figure}
\makeatother
\begin{center}
\subfigure[TC:0000400 Word Cloud]{\includegraphics[width=100px]{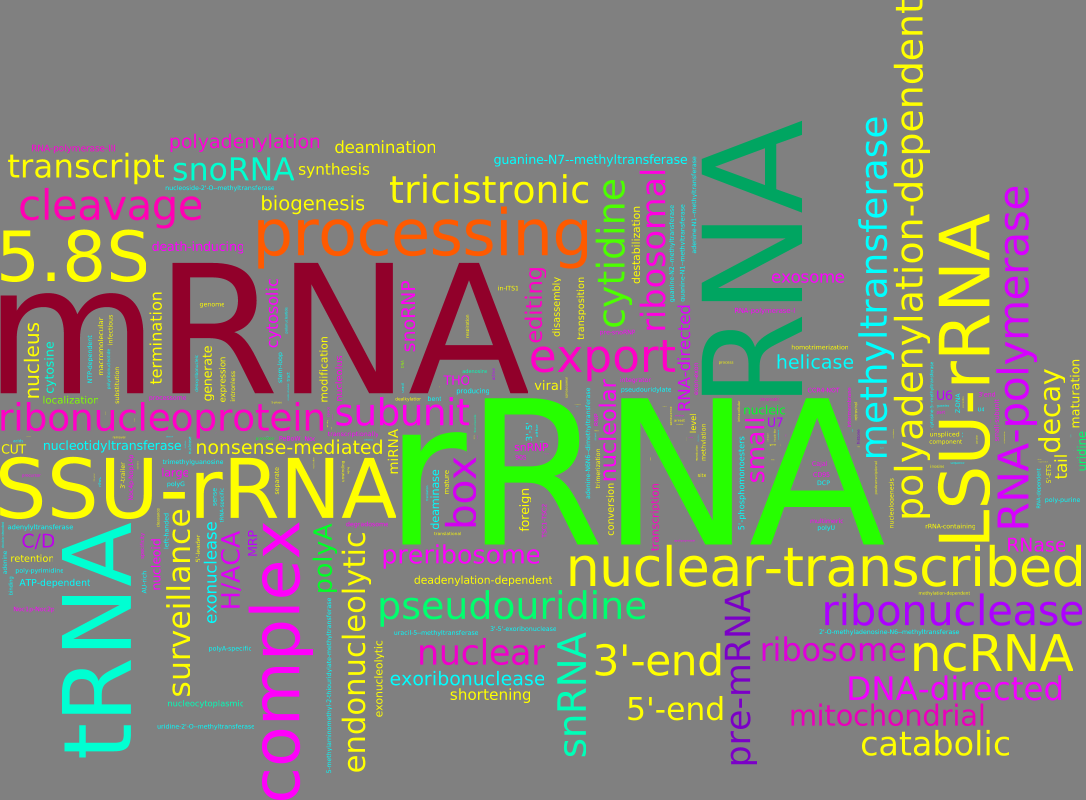}\label{Comm400Cloud}}
\hspace{10px}
\subfigure[TC:0000061 Word Cloud]{\includegraphics[width=100px]{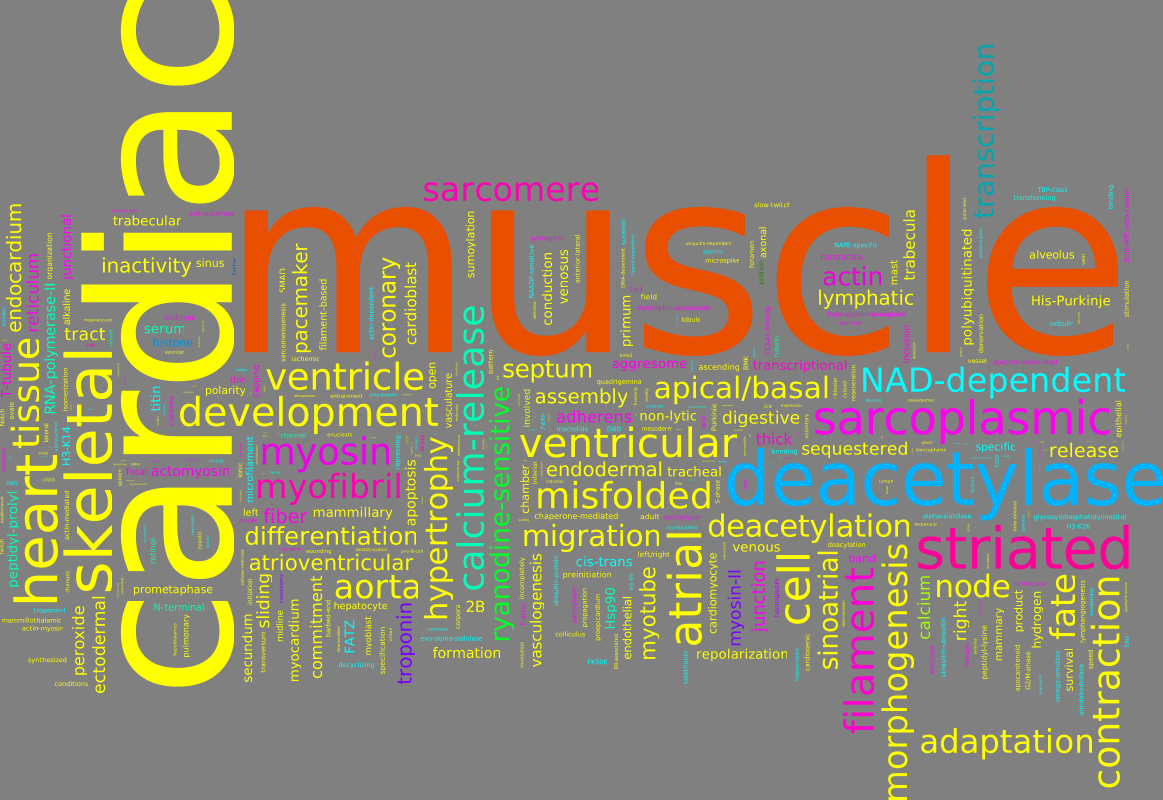}\label{Comm61Cloud}}
\subfigure[GO:0000003 Word Cloud]{\includegraphics[width=100px]{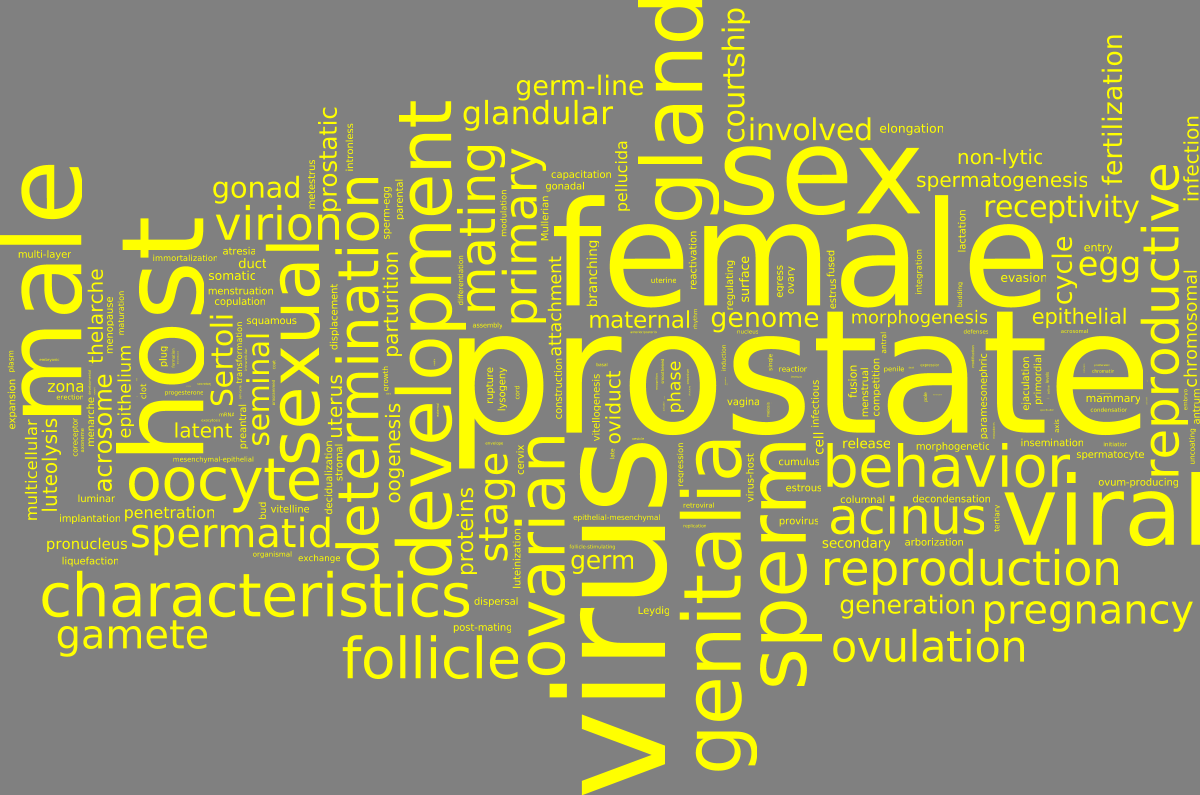}\label{Branch1Cloud}}
\hspace{10px}
\subfigure[GO:0002376 Word Cloud]{\includegraphics[width=100px]{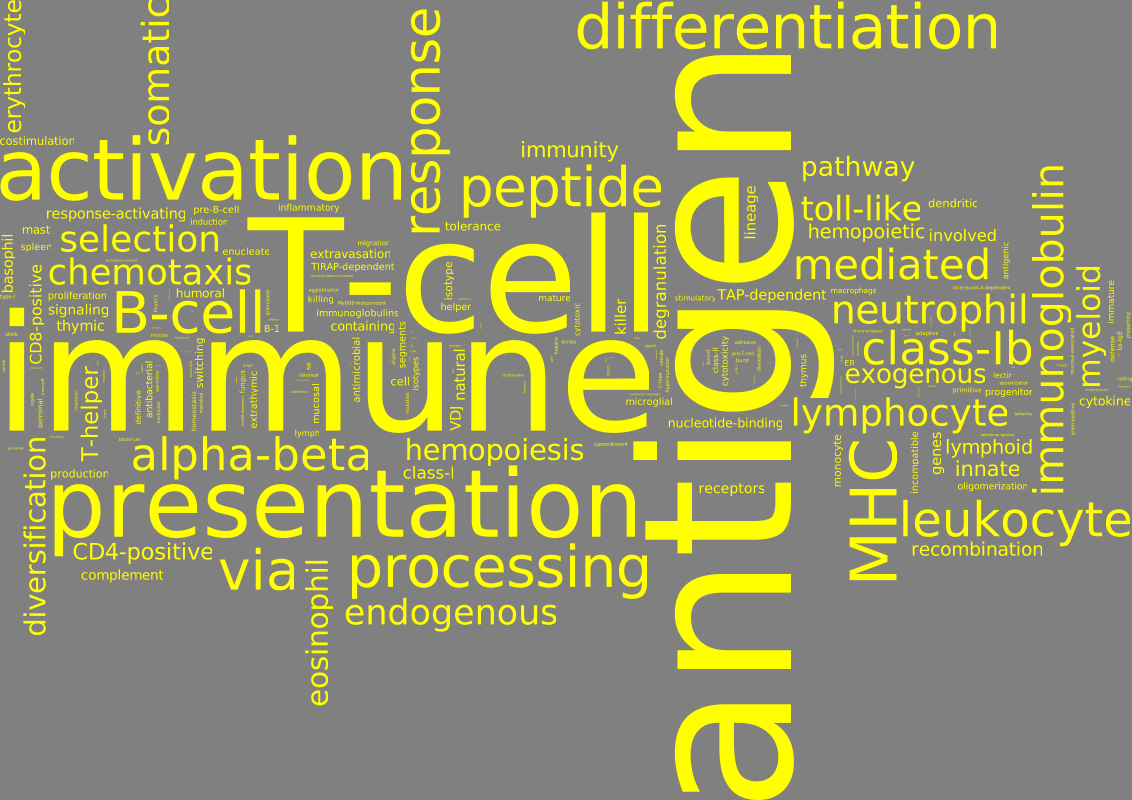}\label{Branch743Cloud}}
\subfigure[Annotation Enrichment Analysis]{\includegraphics[width=240px]{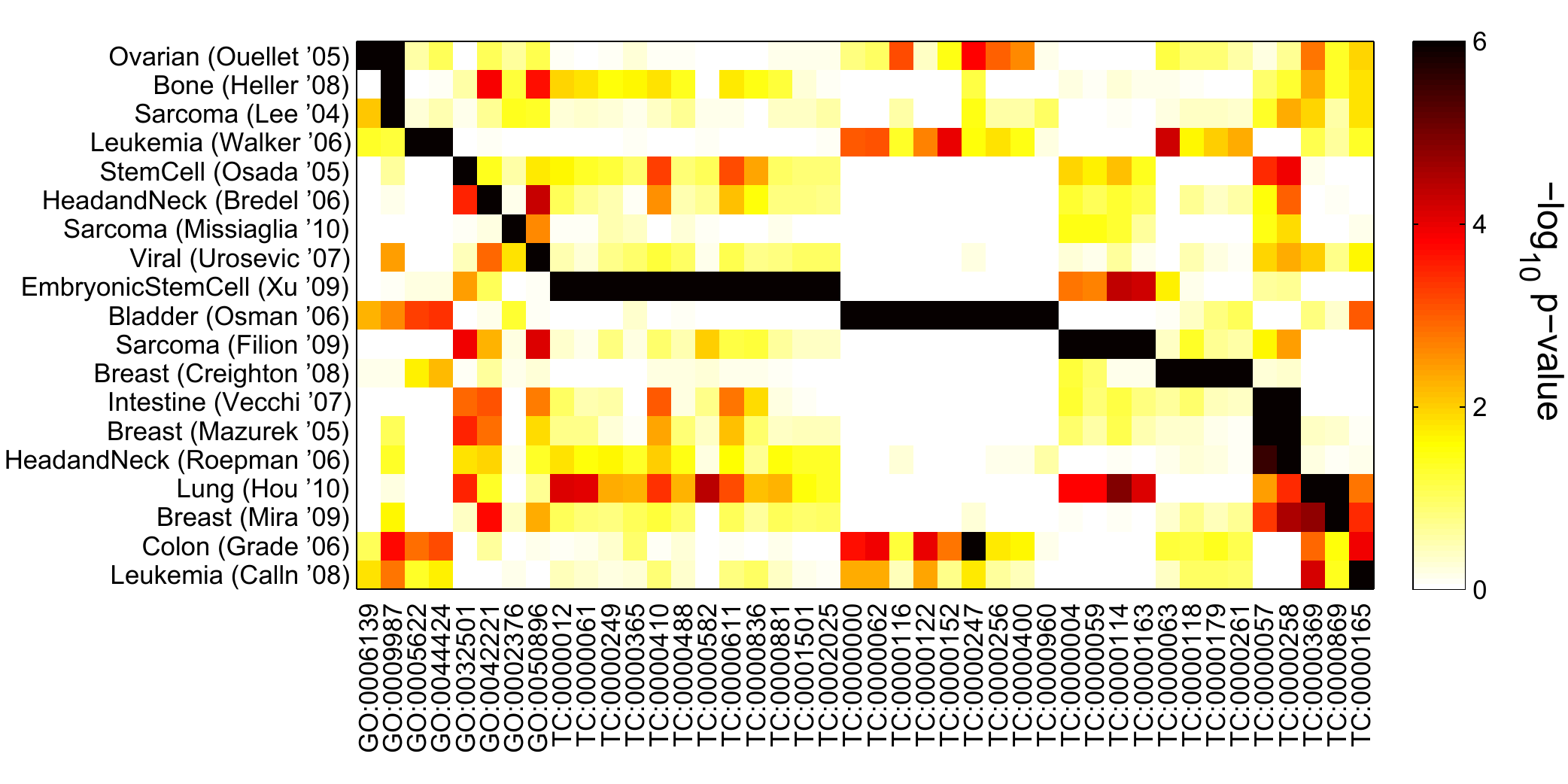}\label{GOheatmap}}
\end{center}
\caption{(A-D) Term Communities (TC:0000400, TC:0000061) and branches (GO:0009607, GO:0050896) summarized as word clouds.  In each case the color of a word represents how often the term description containing that word belongs to each of the primary domains (BP:yellow, MF:cyan, CC:magenta, also see Figure \ref{CommunityDAG} for mixed-domain coloration) and size represents that word's statistical enrichment in that community/branch. (E) A heat map showing the statistical enrichment of selected cancer signatures (see text) in GO branches and term communities.}
\label{CommWordCloud}
\end{figure}

In general, we observe that term communities contain slightly more statistically enriched associations with these experimental signatures than GO branches (which are widely used in functional enrichment analysis) and we verified that this level of enrichment is absent for randomly constructed communities (see Supplemental Figure \ref{CompareEnrichment}).  Knowing that our communities are statistically associated with experimental gene signatures, we next sought to know if there was a context in which our term communities captured biological information from these signatures that is missed by the branches, or vice versus.  Thus we selected gene signatures that were significantly enriched ($p<10^{-6}$) in at least one community/branch  but not significantly enriched ($p>5\times10^{-5}$) in any GO branch/community, respectively.  Figure \ref{GOheatmap} shows a heat map of the enrichment values for the nineteen signatures that met this criteria across any community or branch statistically enriched in at least one of those signatures.

It is immediately striking that of these signatures, the majority are enriched in communities and not GO branches.  Two signatures, in particular, are enriched in a collection of communities.  The first, an embryonic stem cell signature \cite{Xu09}, represents genes that are up-regulated in cardiomyocytes compared to non-selected embryoid bodies and hESC. The communities represented in this signature contain several different themes, all consistent with the expected properties of genes selected from stem cells and related to the heart. The corresponding clouds emphasize words such as ``cardiac'' and ``muscle'' (TC:0000012), ``actin'', ``myosin'', and ``filament'' (TC:0000249), ``morphogenesis'' and ``development'' (TC:0000365), ``blood'', ``pressure'' and ``contraction'' (TC:0000582), with the other clouds generally containing these words in different combinations (see for example TC:0000061, illustrated in Figure \ref{Comm61Cloud}).  The second signature is a list of bladder cancer specific genes \cite{Osman06}.  Most of the words emphasized by the community clouds are related to cell proliferation.  For example, it is enriched in TC:0000400 (illustrated in Figure \ref{Comm400Cloud}) and the other clouds emphasize words such as ``cell-cycle'', ``mitotic'', ``meiotic'', ``checkpoint'', ``repair'',  ``replication'', ``recombination'', ``telomere'', ``spindle'', ``complex'', ``DNA'', ``chromosome'', ``histone'', and ``methylation''.  Although the connection to the bladder is not obvious, the connection to cancer and the high rate of cell proliferation in tumor cells \cite{CancerMedicine} is apparent.

\section{Conclusion}

The network structure of gene annotations made to the Gene Ontology has not previously been exploited in a manner that reveals an organization of biological function unique from the published hierarchical classification of the GO DAG.  By analyzing functional annotation data we were able to construct an alternate, natural, and biologically-relevant way in which to categorize cellular functions.  This categorization is structurally and conceptually distinct from the GO DAG and allows us to uncover multiple, strong connections between terms that do not share a parent/child relationship.  It takes advantage of a large amount of data from a variety of sources and creates a classification scheme that is motivated primarily by the data reported rather than the organization of human conceptions.

The term communities defined in this work represent an integration of information across all three primary domains in GO that, to the authors' knowledge, has not previously been investigated systematically.  Using the simple principle of co-annotation we suggest that in the future biological concepts from other classification databases could also be analyzed or even combined with these results.  We concede that the communities defined here likely do not represent the only way to group functional terms outside of the ontology structure.  Even so, we believe that our functional enrichment analysis demonstrates that these term communities, in particular, are more than a mathematical phenomenon and have a high potential to be used to better interpret biological data.

%
%

\section*{Acknowledgements}
We wish to thank Geet Duggal for supplying an implementation of the Fast Greedy Community Structure algorithm that included the resolution parameter.

%
%

\clearpage

\setcounter{figure}{0} \renewcommand{\thefigure}{S\arabic{figure}}
\setcounter{table}{0} \renewcommand{\thetable}{S\arabic{table}}

\section*{\large Supplemental Material}

In this section we provide additional materials meant to compliment and expand upon the analysis in ``Finding New Order in Biological Functions from the Network Structure of Gene Annotations.''  In the first section, ``Communities Found Across Multiple Resolutions'', we provide more detail about how we used the resolution parameter to generate multiple, overlapping, yet unique communities of terms.  In the second section, ``Domain-specific Analysis'' we provide analysis of communities of terms compared to branches within each primary domain of GO.  Next, in ``Visualizing GO Branches'' we illustrate relationships between GO branches at different ``levels'' of the GO DAG.  In the section entitled ``Functional Enrichment Analysis'', we briefly explain the methodology used to evaluate the statistical enrichment of gene sets in the term communities and GO branches, and provide analysis showing that this enrichment is absent for randomly generated term communities or randomly generated sets of genes.  Finally, in ``Comparative Species Analysis'' we construct and analyze annotation-driven term networks using gene annotations for sixteen additional organisms and compare those networks with each other as well as the human results presented in the main text.

For information regarding the methodology used to illustrate the term communities or generate the word clouds in the main text, see ``Supplemental Methods'' below.

\subsection{Communities Found Across Multiple Resolutions}
\label{CollapseCommunities}

In order to find additional viable partitions of the annotation-driven term network, representing different resolutions, we varied the weighting parameter (see $r$ in Equation 3, and ``Methods'' in the main text).  In addition to the fundamental partition ($r=0$) we chose values of $r$ in geometrically increasing steps ranging from $r=2^{-4}=0.25$ to $r=2^{10}=1024$ such that the final distribution of community sizes would resemble that of the branches.  This procedure resulted in fourteen different partitions of the terms in the annotation-driven network such that, initially, each term can be assigned to exactly fourteen communities, one at each resolution.  We emphasize that while communities at any given resolution do not overlap, communities at different resolutions can be highly overlapping.

\begin{table}[h]
\begin{tabular}{c c c}
Value of r & Number of & Number of New \\
(Resolution) & Communities & Communities \\
\hline
0 & 56 & 56 \\
0.25 & 80 & 51 \\
0.5 & 89 & 44 \\
1 & 116 & 67 \\
2 & 146 & 95 \\
4 & 193 & 148 \\
8 & 320 & 257 \\
16 & 576 & 422 \\
32 & 1007 & 703 \\
64 & 1557 & 965 \\
128 & 2409 & 1439 \\
256 & 3585 & 1965 \\
512 & 4983 & 2375 \\
1024 & 6730 & 2904 \\
\end{tabular}
\caption{The number of communities found at each value of resolution used.  As the resolution is increased, the number of communities found at that resolution increases as well.}
\label{Table1}
\end{table}

\begin{figure*}
\subfigure[Distribution of Branch and Community Sizes]{\includegraphics[width=120px]{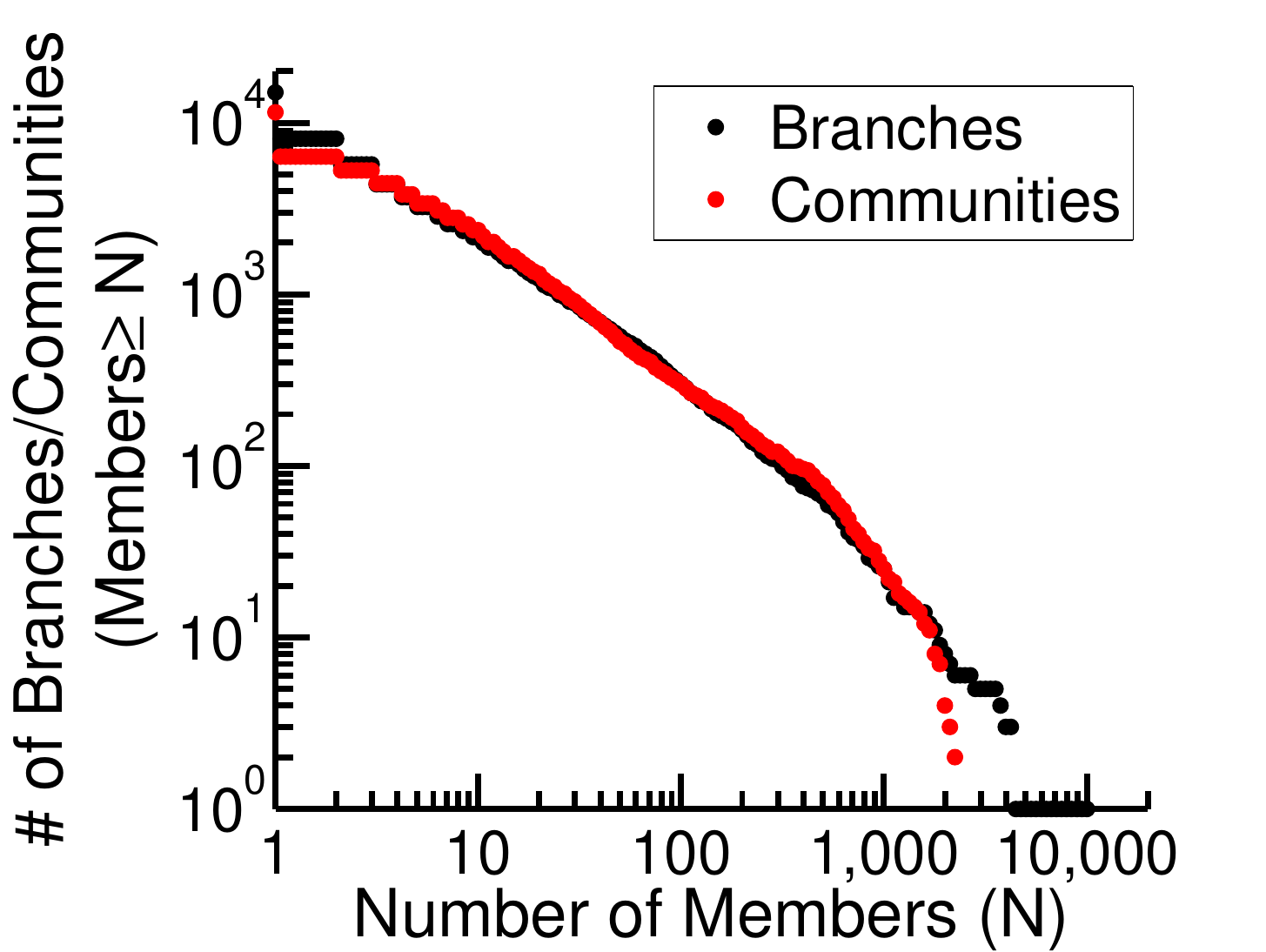}\label{BranchCommSizeDist}}
\subfigure[Biological Process Terms]{\includegraphics[width=120px]{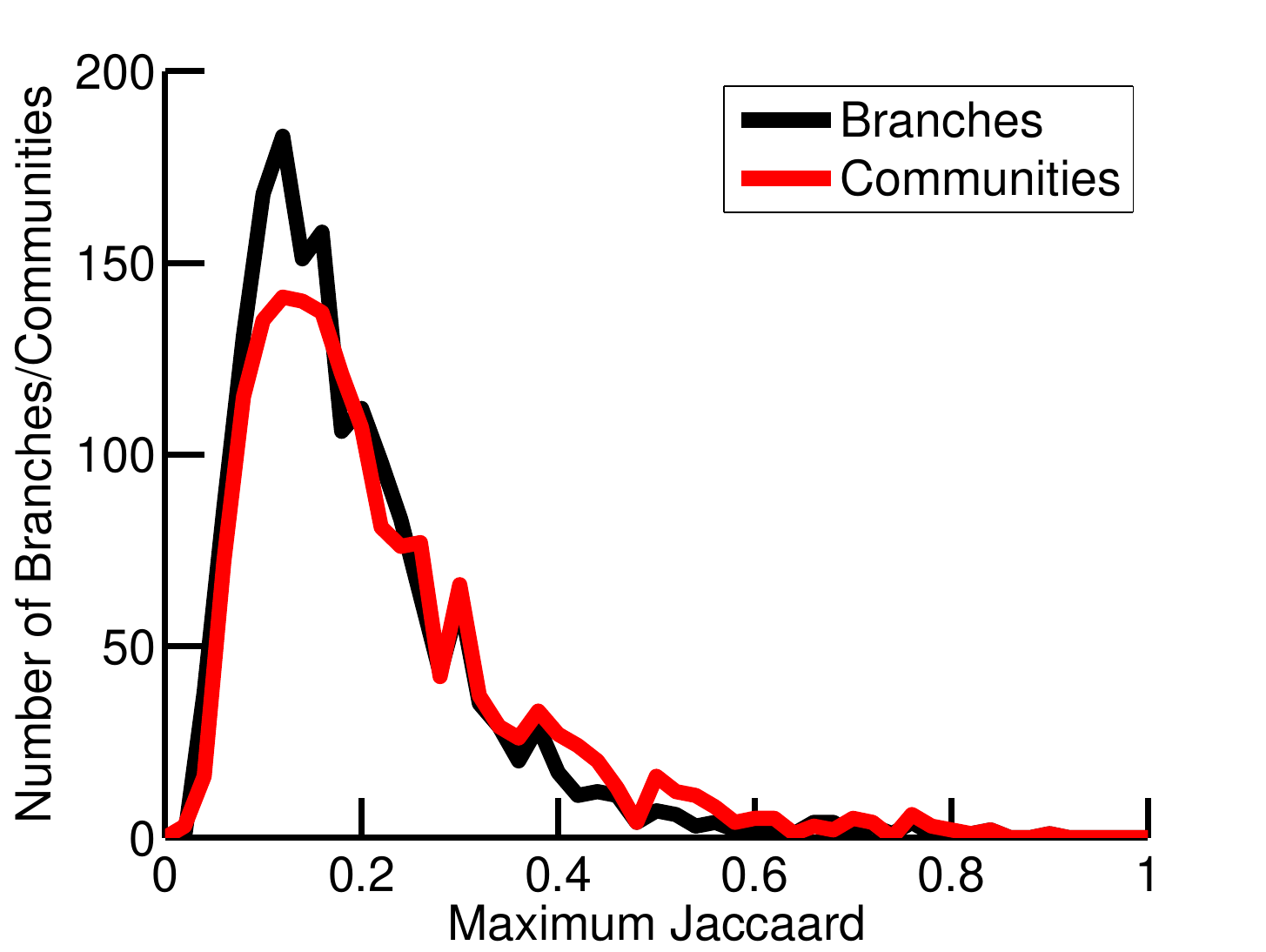}\label{MaxJBP}}
\subfigure[Molecular Function Terms]{\includegraphics[width=120px]{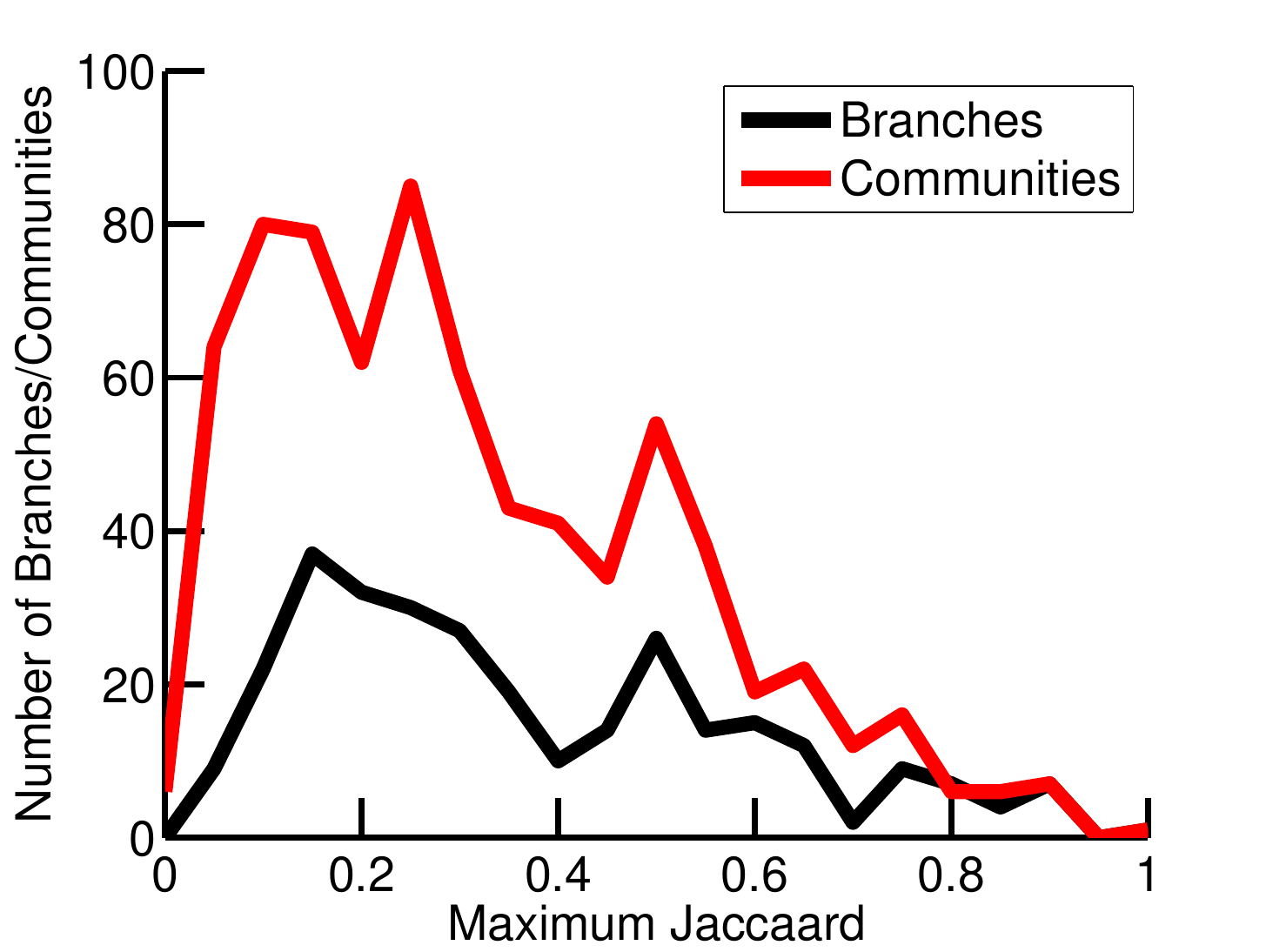}\label{MaxJMF}}
\subfigure[Cellular Component Terms]{\includegraphics[width=120px]{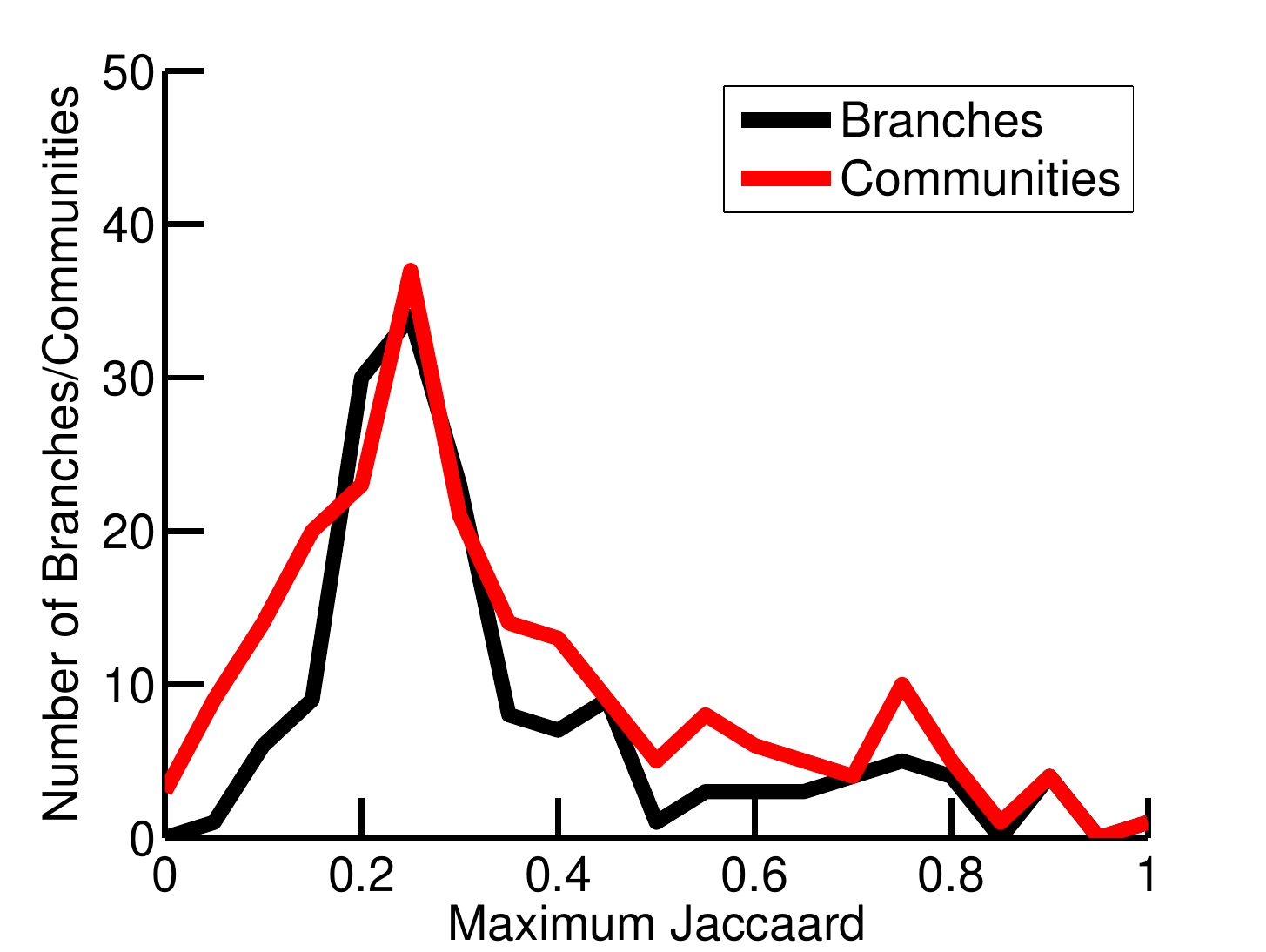}\label{MaxJCC}}
\caption{(A) The cumulative Distribution for the sizes of all branches in the Gene Ontology and all unique term communities found at the various resolutions.  (B-D) Distribution of $J_m$, the maximum similarity a domain-specific community or branch with ten or more members has compared to all other domain-specific branches or communities with ten or more members, respectively.  Although a small number of communities and branches have similar memberships, most are very dissimilar.}
\label{MaxJaccardDomain}
\end{figure*}

We point out that it is possible for the membership of a community found at one resolution to be identical to the membership of a community found at another resolution.  In order to eliminate completely redundant community information from our found communities, at each resolution, we determined if any of the communities found at that resolution were identical in membership as a community found at a lower resolution.  If so, we ``collapsed'' the two community assignments into that of the community from the lower resolution.  For example, of the $80$ communities found at a resolution value of $r=0.25$, $29$ had an identical membership to one of the $56$ communities found at $r=0$.  We removed those $29$ ``redundant'' communities from the $r=0.25$ partitioning, retaining only their $r=0$ assignment, and record that at $r=0.25$ only $51$ additional communities are found.  Similarly, of the $89$ communities found at a resolution of $r=0.5$, $45$ of them are identical in membership to one of the $107$ unique communities found at $r=0$ or $r=0.25$.  These ``redundant'' communities were removed and we record only $44$ additional communities found at $r=0.5$. This procedure was repeated for the remaining resolutions, resulting in 11491 ``unique'' communities (see Table \ref{Table1}).  The cumulative distribution of the number of members in these communities is shown in Figure S\ref{BranchCommSizeDist}.  For comparison, on the same graph we also plot the cumulative distribution of the number of annotations made to GO branches.  These distributions are heavy-tailed and demonstrate that the number and sizes of branches and communities are similar.  The heavy-tailed distribution for branches is a result of the hierarchical DAG structure as the members of each branch are also members of their parent branch(es) (see \cite{Glass12, citeulike:11319934}).

\subsection{Domain-specific Analysis}

The Gene Ontology is broken into three, fully independent, primary domains, each of which takes the form of a directed acyclic graph \cite{GO2001}.  In the main text we combined information from all three primary domains to construct the annotation-driven term network.  Because of this, the dissimilarity between the GO branches and the term communities we find by partitioning this network may be at least partially attributable to the fact that members of a particular community can belong to any of the primary domains whereas members of a particular GO branch must all belong to the same primary domain based on the construction of the hierarchy.  To address the extent of this issue, we constructed three ``domain-specific'' term networks by using only terms specific to a particular primary domain and the gene annotations made to those specific terms.  We partitioned each of these networks using the same resolution values as were used previously and, as with the partitions using all annotations, we retained only the ``unique'' set of communities for the following analysis (see Section \ref{CollapseCommunities}).  Table \ref{Table2} shows the number of communities found and the number of branches defined in GO for each of the primary domains.

\begin{table}[b]
\begin{tabular}{c | c | c |}
 & Number of & Number of \\
 & Branches & Communities \\
\hline
All Terms & 15033 & 11491 \\
\hline
BP Terms & 10192 & 9043 \\
\hline
MF terms & 3634 & 5423 \\
\hline
CC Terms & 1207 & 2107 \\
\hline
\end{tabular}
\caption{The number of branches defined within each primary domain as well as the number of communities found by varying the resolution parameter and partitioning a term-network derived by gene annotations made to terms in this primary domain.  The large size of the ``Biological Process'' primary domain compared to the others is evident.}
\label{Table2}
\end{table}

Next we compared the membership of these communities and branches using the Jaccard similarity (see ``Term Communities and GO Branches Represent Distinct Collections of Biological Functions'' in the main text).  We did the comparison three times, each time confining the comparison to those branches defined by a particular primary domain and the term communities derived from the network constructed using those terms and their corresponding gene annotations.  The distribution of $J_m$ (Figure \ref{MaxJaccardDomain}), the maximum similarity a community has to any branch, or vice versus, for each of the domains, is not strikingly different from that presented when using information collected from all three domains (see Figure \ref{BranchCommJaccardDist} in the main text).  There might be slightly more similarity when directly comparing communities and branches derived from either the ``Molecular Function'', or ``Cellular Component'' primary domains, but we remind the reader that the majority (approximately two-thirds) of the terms in GO belong to the ``Biological Process'' primary domain, and this distribution (Figure S\ref{MaxJBP}) is practically indistinguishable from the one presented in Figure \ref{BranchCommJaccardDist} of the main text.

\subsection{Visualizing GO Branches}

To better understand the relationships between the branches in the Gene Ontology, we visualized branches in a manner similar to the way we visualized the relationships between term communities found at different resolutions (see Figure \ref{CommunityDAG} in the main text).  First, we determined a ``level'' for each GO branch in order to segregate the branches in a manner similar to the resolution parameter.  To begin, the head nodes of the three primary domains (``Biological Process'', ``Molecular Function'', and ``Cellular Component'') were assigned to level one.  Next, we determined branches for which the head-node is a term that has a parent-child relationship with only one of these three level-one terms, and assigned those terms a level of two.  Continuing, head-nodes that have parent-child relationships with only level-one or level-two terms were given a level assignment of three, and so on.  This assignment was repeated until all head-nodes were assigned a ``level''.  Branch levels were then defined based on the level of its head-node.

\begin{figure*}
\includegraphics[width=500px]{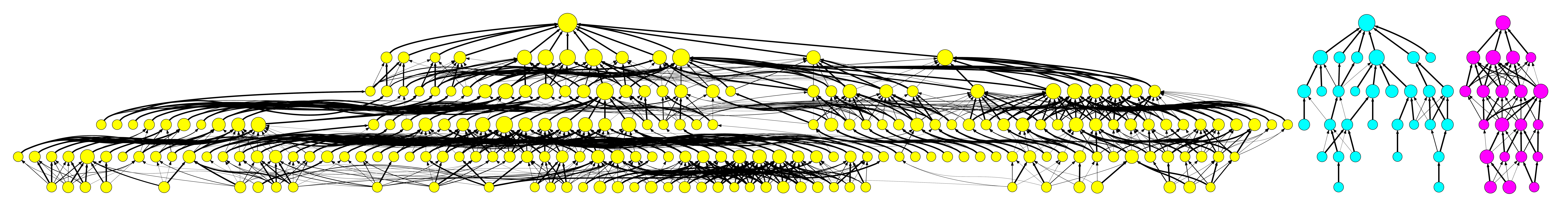}
\caption{Visualization of branches of GO terms.  Each branch is represented as a circle whose radius scales as the log of the number of terms in the branch.  The width of the line connecting two branches is proportional to the percentage of terms in the child branch that are also in the parent branch.  Color represents whether the terms in the branch belong to the BP (yellow), MF (cyan) or CC (magenta) primary domain.}
\label{VizBranches}
\end{figure*}

We visualized branches with one-hundred or more term-members at the six lowest levels (Figure \ref{VizBranches}), lining up branches with the same level assignment horizontally.  Each branch is represented by a single circle, whose radius scales as the log of the number of terms belonging to that branch and whose color represents whether the terms in the branch belong to the BP (yellow), MF (cyan) or CC (magenta) primary domain.  Between the branches found at adjacent levels, we draw a line from a branch at a higher level to a branch at a lower level if at least 10\% of the members of the branch from the higher level also belong to the branch at the lower level.  The thickness of the line is indicative of the overlap in membership between the two branches.

Unsurprisingly, we observe three distinct sets of inter-connected branches corresponding to branches belonging to each of the three primary domains.  Like in the visual representation of the term communities across different resolutions (see Figure \ref{CommunityDAG} in the main text), we observe a lot of ``cross-talk'' between branches at adjacent levels, whereby a branch at a given level is very likely to contain members from multiple branches at a lower level.  We note that in this representation, an individual term can be a member of multiple branches at the same ``level''.  This is in contrast to segregating communities by resolution, in which case each term only appears once on a given resolution-row.  As a consequence, the inter-level connections between branches are somewhat structurally different from connections between communities found at adjacent resolutions.  Namely, branches that share a term will necessarily also share a set of parent branches.  Redundancy of the same term member(s) across multiple branches at a given level is visually evident among ``Biological Process'' (yellow) branches, where there exists groups of branches at each level that connect primarily to the same set of branches at a lower level.

\subsection{Functional Enrichment Analysis}

We also wanted to test how our communities might be used in functional enrichment analysis, a very common application of the Gene Ontology.  Traditionally, each branch of GO is collapsed to its head node and all the genes annotated to that head node are grouped into one ``set.'' (Note that this set is the same as the genes annotated to the entire branch because of the propagation of gene annotation assignment, see the ``Introduction'' in the main text).  Recently there has been evidence that this approach over-simplifies the complex structure of the Gene Ontology and has the potential to mis-represent the enrichment of gene sets in branches \cite{citeulike:11319934}.  Therefore, we choose to use Annotation Enrichment Analysis (AEA) to evaluate the functional enrichment of experimentally-derived gene sets in both the GO branches and our term communities.  AEA allows the user to specify a collection of terms and a collection of genes and uses a randomization protocol to evaluate the probability that these two sets are more connected than by chance.

\begin{figure*}
\subfigure[Annotation Enrichment Analysis]{\includegraphics[width=500px]{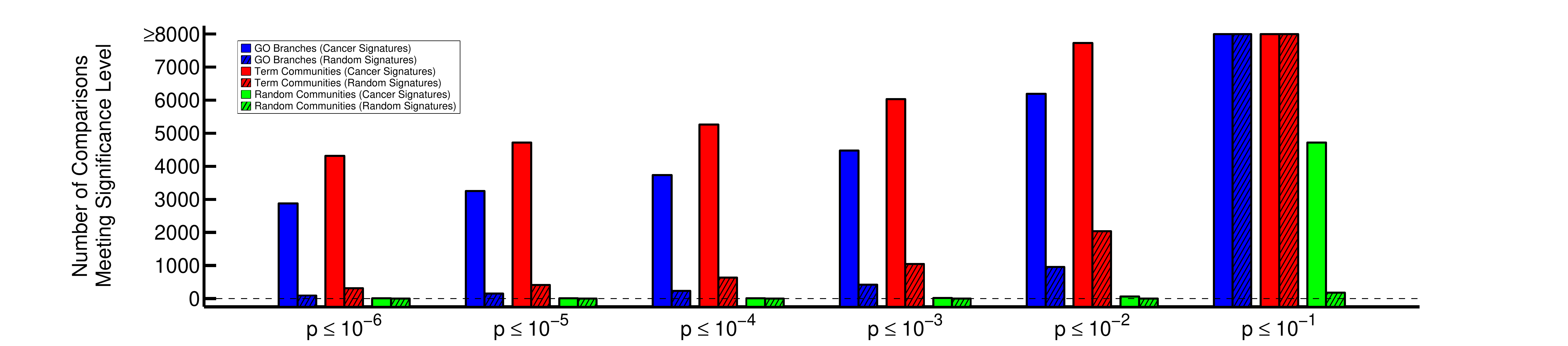}\label{CompareAEA}}
\subfigure[Fisher's Exact Test]{\includegraphics[width=500px]{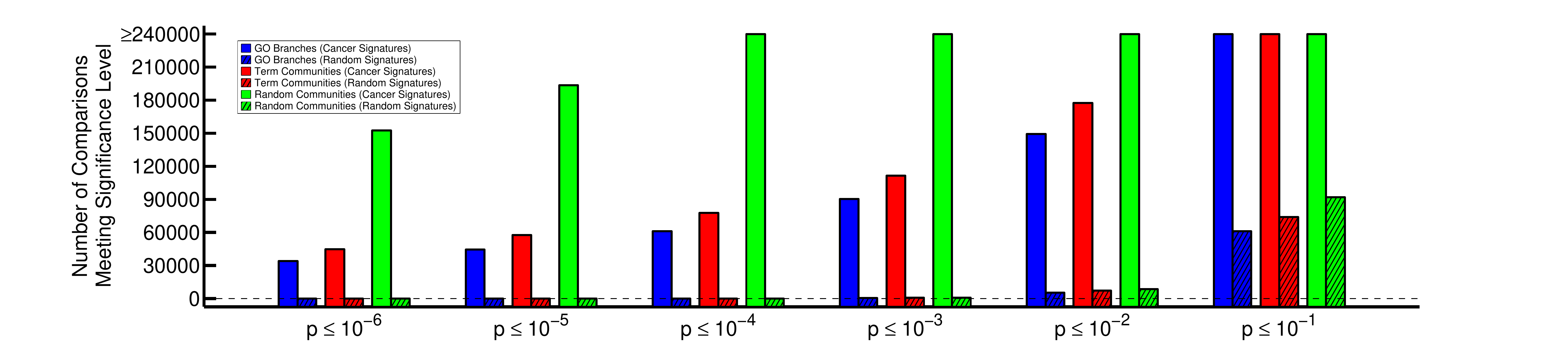}\label{CompareFET}}
\caption{Level of statistical enrichment found when comparing branches, communities, and randomly generated communities with either cancer signatures or randomly generated gene sets.  Only branches and our identified term communities show statistical enrichment in experimentally-derived gene signatures, with slightly more enrichment for the communities compared to the branches.}
\label{CompareEnrichment}
\end{figure*}

We ran AEA using one million randomizations defining collections of genes using a public database of Cancer Signatures \cite{GeneSigDB2012} and using collections of terms defined by (1) membership in the term communities; or (2) membership in GO branches.  We plot the number of pairs of term and gene ``collections'' that are enriched beneath various significance thresholds (Figure S\ref{CompareAEA}) and observe clear statistical enrichment of experimentally-derived gene signatures in both term communities and GO branches with several thousand community-signature and branch-signature pairs enriched at a p-value significance less than $10^{-6}$.

Next, we wished to verify that this enrichment was not an artifact of the community construction -- namely, we wanted to verify that experimental gene signatures are not enriched in random collections of terms and that term communities are not enriched in random collections of genes.  Therefore we constructed ``random'' term communities by taking the community assignments of terms and swapping term labels.  Similarly, we constructed ``random'' gene sets by randomly swapping gene labels.  This gives us random term communities and random gene sets with both the same size and relative overlap as the real term communities and the experimentally-defined signatures.  We then ran AEA an additional four times, using: (1) random communities and the experimentally-defined signatures; (2) term communities and random gene sets; (3) branches and random gene sets; and (4) random communities and random gene sets.  The result of AEA indicated no enrichment using either random term communities or random sets of genes (Figure S\ref{CompareAEA}), showing that the term communities generated by partitioning the annotation-driven term network contain useful biological information.

AEA evaluates the overlap in annotations made by a set of genes or to a set of terms.  Other, traditional functional enrichment analysis procedures, however, often use Fisher's Exact Test (FET) to evaluate the overlap in two sets of genes -- one defined by a gene set or signature of interest, and the other defined by taking the set of genes annotated to all the terms in a GO branch (same as the genes annotated to the parent node).  Although there is evidence that such analysis is highly sensitive to the annotation degree of genes and terms \cite{citeulike:11319934}, we wanted to see if our communities were enriched in cancer signatures using this more traditional approach.  Therefore, for each GO branch, term community and random community, we took the collection of genes annotated to all terms in that branch/community/random community, and assigned this set of genes to represent that branch/community/random-community.  We then evaluated the significance of overlap between theses sets of genes and sets of genes as defined by cancer signatures, or random sets of genes.

As with AEA, both term communities and GO branches show enrichment in experimentally-defined gene signatures using FET, with term communities perhaps having a slightly greater level of enrichment (Figure S\ref{CompareFET}).  At first it is surprising to observe that \emph{random communities} also show a large amount of enrichment in the experimental gene signatures -- much more than either the branches or real communities!!  In retrospect, however, this serves to highlight a known weakness of FET to incorrectly over-estimate the significance of overlap when genes in a set contain a higher than expected number of annotations.  Whereas our random gene sets represent a random sampling from all genes annotated to GO, the genes collected in the signatures published by GeneSigDB are biased in that they are generally annotated much more frequently to GO than one would expect by chance (see \cite{citeulike:11319934}).  Furthermore, by taking all genes annotated to a collection of terms, highly annotated genes are also more likely to be represented in the gene sets representing the branches, term communities, and the random communities.  The enrichment of the experimental gene signatures in the random communities, therefore, is attributable to the fact that FET finds significant overlap between sets containing an abundance of highly annotated genes, independent of the biological content of those sets.  For more discussion, see \cite{citeulike:11319934}.

\subsection{Comparative Species Analysis}

Even though the Gene Ontology hierarchy establishes a species-independent terminology, one could imagine constructing networks of terms using species-specific gene annotations, and thereby constructing species-specific term communities.  These communities would reflect the biological terms that are performed by the same sets of genes in a particular species and wouldn't necessarily be the same across different species.  In this section we evaluate the similarity between partitions of GO terms derived from the annotations of seventeen model organisms, including thale cress (\emph{Arabidopsis thaliana}), \emph{Escherichia coli}, slime mold (\emph{Dictyostelium discoideum}), \emph{Aspergillus nidulans}, three types of yeast including \emph{Candida albicans}, budding yeast (\emph{Saccharomyces cerevisiae}), and fission yeast (\emph{Schizosaccharomyces pombe}), worm (\emph{Caenorhabditis elegans}), fruit fly (\emph{Drosophila melanogaster}), zebrafish (\emph{Danio rerio}), chicken (\emph{Gallus gallus}), pig (\emph{Sus scrofa}), cow (\emph{Bos taurus}), dog (\emph{Canine lupus familiaris}), mouse (\emph{Mus muculus}), rat (\emph{Rattus norvagicus}) and human (\emph{Homo sapiens}).  We downloaded gene annotation files for each of these species and projected term-term networks (see Section ``Constructing a Term Network from Gene Ontology Annotations'' in the main text).  We then partitioned each of these networks into communities (see Section ``Identifying Communities of GO terms'' in the main text).  For simplicity we choose to focus only on the fundamental partition (resolution parameter $r=0$, see Equation 3 in the main text).  This results in exactly one discreet partitioning of GO terms associated with each species.

\begin{figure}
\includegraphics[width=220px]{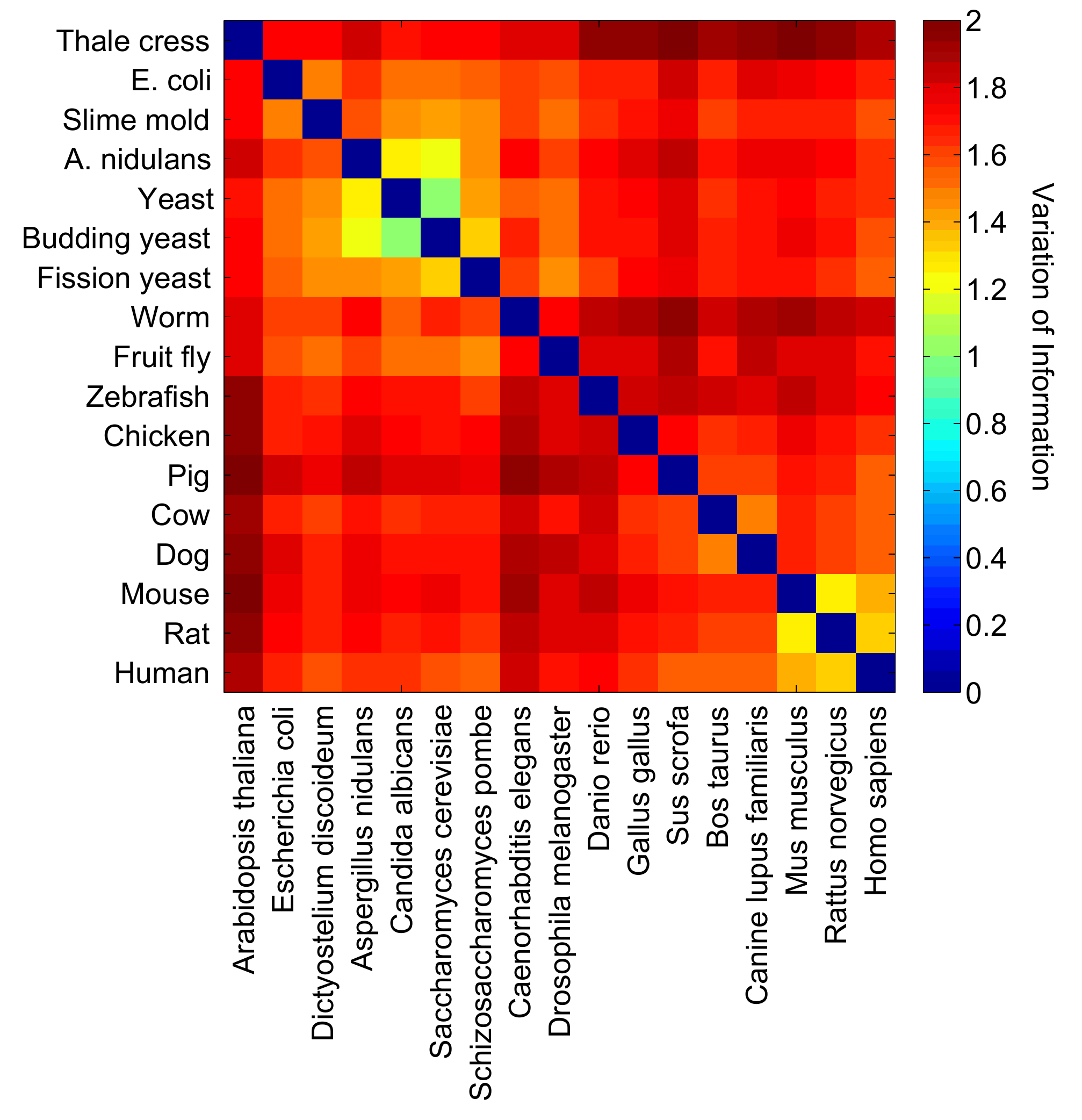}
\caption{The variation of information between term partitions generated from species-specific projected term networks.  Labels along the x-axis are the scientific name for the species and labels along the y-axis are the common name for the species.  Although the partitions are more similar than random ($VI \approx 2.4$), they are still far from being identical ($VI=0$), indicating that these species-specific communities of functional terms carry distinct information.}
\label{CompareSpecies}
\end{figure}

Comparing community structure identification is an ongoing area of research in the complex systems field and there are multiple proposed methods for comparing two discreet partitions of a set of nodes (e.g. \cite{meilaVI,danonNMI}).  One we will employ here is the variation of information ($VI$) \cite{meilaVI}:
\begin{equation}
VI(X,Y)=H(X)+H(Y)-2MI(X,Y)
\label{VariationOfInformation}
\end{equation}
where $H(X)$ is the Shannon's entropy associated with a partition, $X$, and $MI(X,Y)$ is the mutual information between two partitions, $X$ and $Y$. $VI(X,Y)$ represents a distance between the information shared between two partitions, $X$ and $Y$, with $VI(X,Y)=0$ indicating identical partitions.

We calculated the VI between the partitions for every pair of species, using only terms common to both partitions when different sets of terms are annotated in the different species.  Figure \ref{CompareSpecies} shows these values.  We also calculated a VI value for a random shuffling of community assignments in each pair of species and observe that ``random'' VI takes a value of approximately $2.4\pm0.12$.  The actual VI is generally lower than this random value showing that there is some shared information; however, most comparisons are much closer to this random value than to ``perfect'' agreement ($VI=0$), indicating that these term partitions are far from identical.

Interestingly, there is a relatively higher level of similarity between the species belonging to the Fungi Kingdom (\emph{A.nidulans}, yeast, budding yeast and fission yeast).  On the other hand there is higher dissimilarity between animals belonging to the Chordata phylum (zebrafish, chicken, pig, cow, dog, mouse, rat, human), both between each other and compared to the other species.  This could be a consequence of evolutionary diversity playing a more dominant role in the organization of biological function in these organisms, perhaps through more complex regulatory mechanisms such as epigenetics.  The exception is when comparing mouse, rat and human, which is not entirely surprising given the extent to which mouse is used to mimic the human system in laboratory experiments.

Although some of the differences between the species-specific term communities may be due to variations in the annotation practices among groups that supply annotations to GO, it is also likely that they reflect real, biological differences in the cellular organization of these systems.  We suggest that using these partitions of terms in a species-specific context may enhance the results of functional analysis for these model organisms.  Furthermore, identifying the exact differences between these communities may uncover important cellular properties of various species, an investigation we leave to future work.

\section*{\large Supplemental Methods}

In this section we provide additional information on the methodology used to illustrate the term communities across different resolutions (Figure \ref{CommunityDAG} in the main text) and generate the word clouds representing the biological content of those communities (Figure \ref{CommWordCloud} in the main text).

\subsection{Illustrating Community Structure at different resolutions}

To better understand the relationships between the communities found at different resolutions, we visualized the uniquely-found term communities with ten or more members for the six lowest values of resolution used ($r=\{0,0.25,0.5,1,2,4\}$) (Figure \ref{CommunityDAG} in the main text).  We line up the communities found at each resolution and visualize each as a circle whose radius scales as the log of the number of term members found in that community and whose color corresponds to the percentage of ``Biological Process'', ``Molecular Function'' and ``Cellular Component'' terms that belong to that community.  In other words, for each community we count the number of members in that community from the ``Biological Process'' domain and divide by the number of members in the entire ``Biological Process'' domain.  We then do the same things for the other two domains.  After these percentages are calculated, within each community they are ``normalized'' by dividing by the maximum found percentage such that the vector representing that community's domain content has at least one member with a value of one.  We can think of these values in terms of a three-part cmy color vector.  The normalization process causes communities with an equal percentage from all three primary domains to be colored black (cmy color vector equal to [1,1,1]), and those with members only from one primary domain to be exactly yellow ([0,0,1]) for ``Biological Process'', cyan ([1,0,0]) for ``Molecular Function'', or magenta ([0,1,0]) for ``Cellular Component''.

Between the communities found at adjacent resolutions, we draw a line from a community at a higher resolution to a community at a lower resolution if at least 10\% of the members of the community from the higher resolution also belong to the community at the lower resolution.  Line thickness scales linearly based on the percentage of members of the community from the higher resolution that belong to the community at the lower resolution.  Note that although connections are only made between communities in adjacent resolutions, sometimes the parent community is identical to another community found at an even lower resolution, in which case the connection is made from the child community to the copy of the parent at its lowest found resolution.

\subsection{Capturing Biological Information In Word Clouds}

In order to easily interpret the contents of our communities we summarize the information contained in each in the form of word clouds using a free word-cloud making program \cite{IBMWordCloud}.  This program automatically configures the orientation of the words in the clouds, but we manually assign each word a relative size and color to represent that word's statistical enrichment in the community and the primary domain that word is representing in the community, respectively.

To begin, for each community, we determine all the descriptions corresponding to the member terms of that community and count the number of times an individual word appears across all these descriptions.  Then, for each of these words, we calculate the statistical enrichment (p-value) of the frequency of that word in the community compared to its frequency across the descriptions of all GO terms using the hypergeometric probability:
\begin{equation}
p = P(N \geq N_{wc} | N_w, N_c, N_{tot}) = \displaystyle \sum_{i=N_{wc}}^{min[N_{w},N_c]} \frac{{N_c \choose i} {{N_{tot}-N_c} \choose {N_w-i}} }{{N_{tot} \choose N_w}},
\label{Hygeo}
\end{equation}
where $N_{wc}$ is the number of times that word appears in the term descriptions specific to a community, $N_w$ is the number of times that word appears across all term descriptions, $N_c$ is the number of individual words in the term descriptions specific to a community, and $N_{tot}$ is the number of individual words across all term descriptions.  We scale the sizes of the words in the word cloud as $-log_{10}(p)$ such that words with the lowest probability of being in the community by chance are given the largest size, and those one might expect by chance are given a size close to zero.

We also colored each word based on the percentage of times the terms that word comes from in a community belongs to each of the primary domains.  Specifically, for each instance of a word in the term descriptions, we determine the domain assignment of that term and give that word instance the same domain assignment.  To color the word in a community-specific context, we determine the domain assignments made to all instances of that word in the community.  We count these instances and divide by the domain assignments made to all words.  This will generate a three-part vector representing the percentage of the primary domain represented by the word in the community.  We ``normalize'' this vector by dividing by the maximum found percentage, resulting in a three-part cmy color vector has at least one member with a value of one.  The described normalization causes a word with an equal percentage from all three primary domains to be colored black (cmy color vector equal to [1,1,1]), and those with members only from one primary domain to be exactly yellow ([0,0,1]) for ``Biological Process'', cyan ([1,0,0]) for ``Molecular Function'', or magenta ([0,1,0]) for ``Cellular Component''.

\end{document}